\documentclass[conference]{IEEEtran}
\IEEEoverridecommandlockouts

\usepackage{cite}
\usepackage{amsmath,amssymb,amsfonts}
\usepackage[normalem]{ulem}
\usepackage{algorithm}
\usepackage{algpseudocode}
\usepackage{amsmath}
\usepackage{subfig}

\usepackage{graphicx}
\usepackage{textcomp}
\usepackage{xcolor}
\usepackage{color, xspace}
\usepackage{multirow}

\def\BibTeX{{\rm B\kern-.05em{\sc i\kern-.025em b}\kern-.08em
    T\kern-.1667em\lower.7ex\hbox{E}\kern-.125emX}}
\begin{document}

\title{HeteFedRec: Federated Recommender Systems with Model Heterogeneity}

\author
{
	Wei Yuan{\small$^1$}\hspace*{6pt}
        Liang Qu{\small$^1$}\hspace*{6pt}
        Lizhen Cui{\small$^2$}\hspace*{6pt}
        Yongxin Tong{\small$^3$}\hspace*{6pt}
        Xiaofang Zhou{\small$^4$}\hspace*{6pt}
	Hongzhi Yin{\small$^{1*}$}\thanks{* Corresponding author.}
 \hspace*{10pt}
  \\
	\fontsize{10}{10}\selectfont\itshape $~^1$The University of Queensland,
        \fontsize{10}{10}\selectfont\itshape$~^2$Shandong University,
        \fontsize{10}{10}\selectfont\itshape$~^3$Beihang University\\
        \fontsize{10}{10}\selectfont\itshape$~^4$The Hong Kong University of Science and Technology\\
	\fontsize{9}{9}\selectfont\ttfamily\upshape$~^1$\{w.yuan,l.qu1,h.yin1\}@uq.edu.au,\fontsize{9}{9}\selectfont\ttfamily\upshape$~^2$clz@sdu.edu.cn\\\fontsize{9}{9}\selectfont\ttfamily\upshape$~^3$yxtong@buaa.edu.cn,\fontsize{9}{9}\selectfont\ttfamily\upshape$~^4$zxf@cse.ust.hk
}

\maketitle

\begin{abstract}
Owing to the nature of privacy protection, federated recommender systems (FedRecs) have garnered increasing interest in the realm of on-device recommender systems.
However, most existing FedRecs only allow participating clients to collaboratively train a recommendation model of the same public parameter size. 
Training a model of the same size for all clients can lead to suboptimal performance since clients possess varying resources.
For example, clients with limited training data may prefer to train a smaller recommendation model to avoid excessive data consumption, while clients with sufficient data would benefit from a larger model to achieve higher recommendation accuracy.
To address the above challenge, this paper introduces HeteFedRec, a novel FedRec framework that enables the assignment of personalized model sizes to participants.
Specifically, we present a heterogeneous recommendation model aggregation strategy, including a unified dual-task learning mechanism and a dimensional decorrelation regularization, to allow knowledge aggregation among recommender models of different sizes.
Additionally, a relation-based ensemble knowledge distillation method is proposed to effectively distil knowledge from heterogeneous item embeddings.
Extensive experiments conducted on three real-world recommendation datasets demonstrate the effectiveness and efficiency of HeteFedRec in training federated recommender systems under heterogeneous settings.
\end{abstract}

\section{Introduction}
Recently, there has been a significant surge in demand for recommender systems in various online services~\cite{chen2020try,wei2007survey,yin2016spatio,wu2020mind,hao2021pre,chen2018tada,yin2015joint},
since they can effectively alleviate information overload by actively discovering users' potential interests.
Conventional recommender systems are trained in a central server with collected user private data, which takes risks of data leakage and raises privacy concerns~\cite{zhang2019deep}.
With the growing awareness of user privacy protection and some recently released regulations, such as GDPR\footnote{https://gdpr-info.eu} in the EU, CCPA\footnote{https://oag.ca.gov/privacy/ccpa} in the USA, and PIPL\footnote{https://personalinformationprotectionlaw.com} in China, it is becoming harder and even infeasible for online platforms to learn recommendation models on centrally collected user data~\cite{liang2021fedrec++}.

Federated learning~\cite{mcmahan2017communication}, as a privacy-preserving paradigm, has achieved remarkable success across various scenarios~\cite{li2020review}.
In order to tackle privacy concerns, researchers have sought to harness federated learning for the development of privacy-aware recommender systems
 namely federated recommender systems (FedRecs), where clients\footnote{In this paper, the terms ``client'' and ``user'' are used interchangeably, as each client is solely responsible for one user to ensure privacy protection.} can collaboratively train a global recommendation model on their local devices without sharing their personal data with any other participants.

Ammand et al.~\cite{ammad2019federated} proposed the pioneering federated recommendation framework.
In this framework, a recommender system is divided into public and private parameters. 
Users and a central server collaboratively learn a recommendation model by transmitting and aggregating the public parameters using specific aggregation strategies.
Due to the advantage of privacy protection, 
several extended versions~\cite{muhammad2020fedfast,liang2021fedrec++,wu2021fedgnn,sun2022survey,imran2023refrs} have been developed based on this fundamental framework to enhance the performance of FedRec.

\begin{figure}[t]
  \centering
  \includegraphics[width=1.\columnwidth]{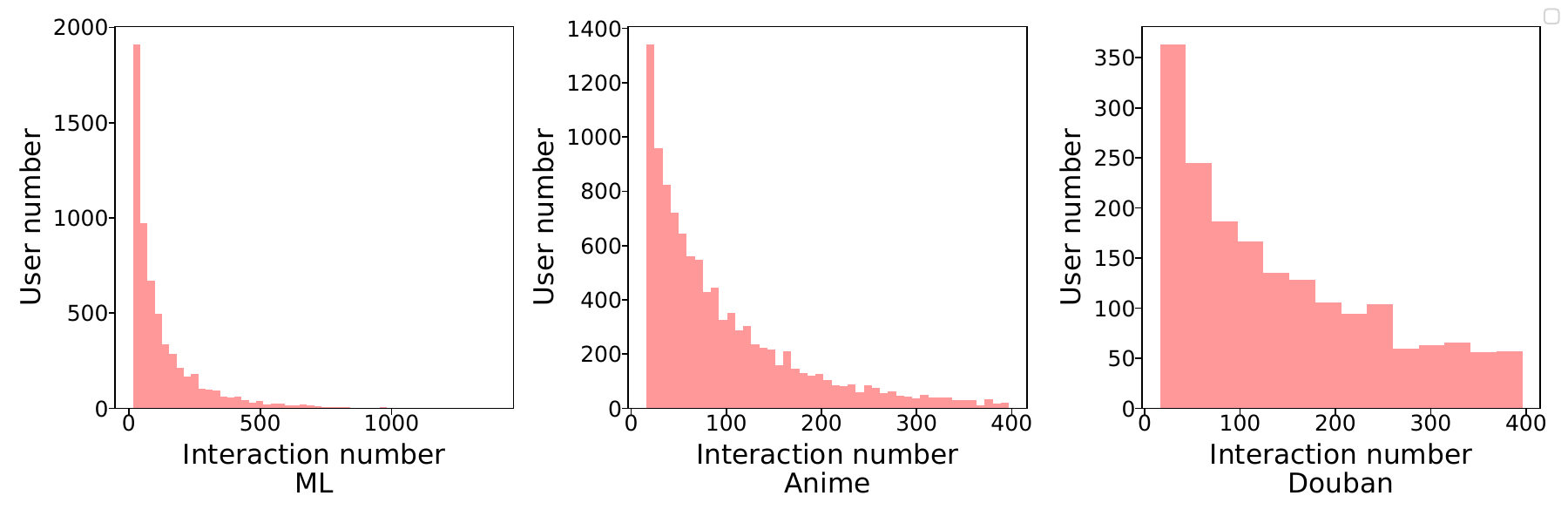} 
  \caption{The distribution of users' item interaction numbers.}
  \label{fig_distribution}
  \vspace{-20pt}
  \end{figure}

  \begin{figure*}[!htbp]
    \centering
    \includegraphics[width=2.\columnwidth]{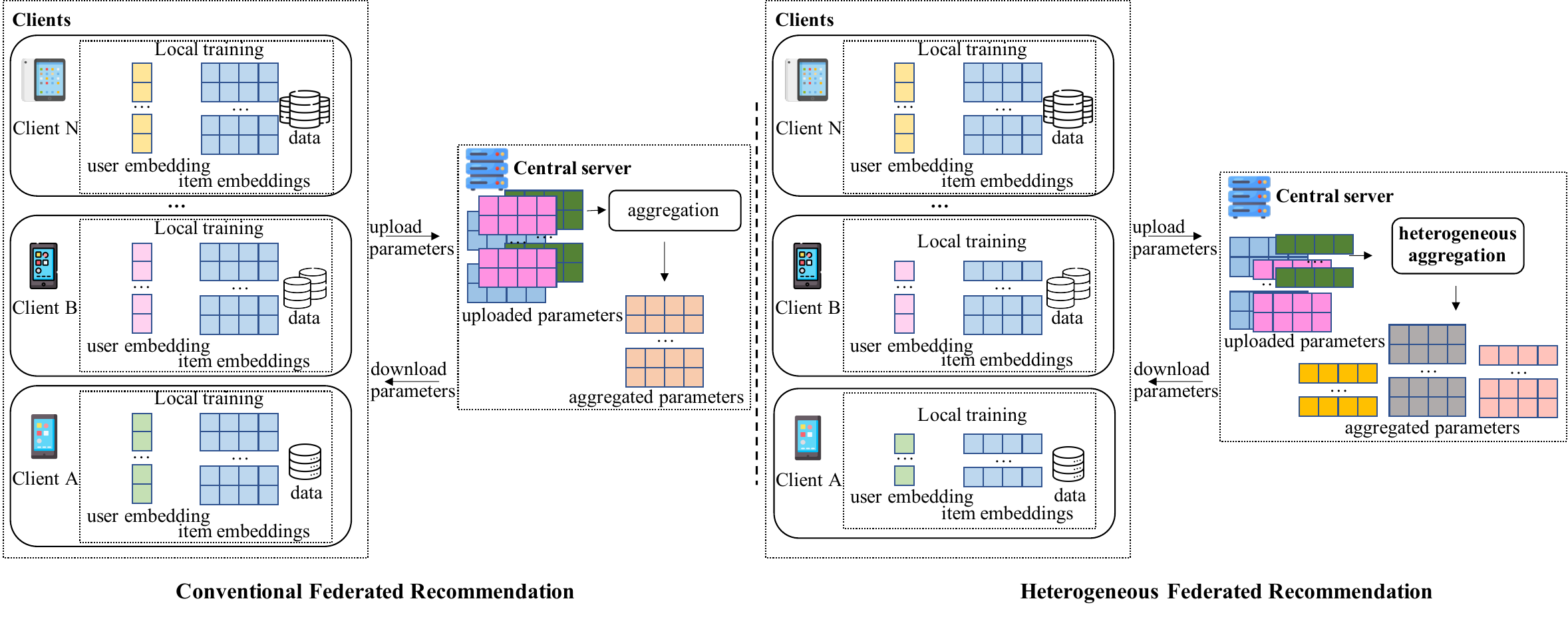} 
    \caption{Conventional Federated Recommender Systems v.s. Heterogeneous Federated Recommender Systems. Heterogeneous federated recommender systems assign public parameters with customized sizes considering clients' data amounts.}
    \label{fig_fedrecvshfedrec}
     \vspace{-10pt}
    \end{figure*}
Although many achievements have been made recently, almost all existing FedRecs operate under the setting where clients collaboratively learn a global recommendation model of the same size since their aggregation strategies (e.g., FedAvg~\cite{mcmahan2017communication}) can only process homogeneous public parameters, as shown in Fig.~\ref{fig_fedrecvshfedrec}.
This setting conflicts with the real applications where clients often possess heterogeneous resources, such as varying amounts of training data\footnote{While this paper emphasizes the necessity of model heterogeneity using data size variance as an example, it's crucial to recognize that such heterogeneity can also tackle some other resource diversity problems, including disparities in computational power, energy constraints, bandwidth, and more.}.
Fig.~\ref{fig_distribution} depicts the user-item interaction number distribution on three recommendation datasets derived from real-world platforms.
The standard deviation of interaction numbers is about $154.2$, $79.8$, and $105.2$ for MovieLens-1M (ML)~\cite{harper2015movielens}, Anime, and Douban datasets, while the average of interaction amounts is $132.8$, $96.1$, and $143.7$, respectively, indicating the substantial difference of data size among clients.
Under this circumstance, training a global recommendation model of the same size on all clients would fall short of achieving the true global optimum~\cite{tan2022towards}.
Clients with limited data would struggle to support the training of the large global model, and their model updates might even have a negative impact on the performance of the global model, as their local models can only be updated for few times by gradient decent~\cite{ruder2016overview} on insufficient data. 
Conversely, clients with abundant data would prefer to train a larger model to obtain more appropriate recommendations. 
Given these limitations, there is an urgent need for a federated recommendation framework that enables the collaborative training of heterogeneous models\footnote{In this paper, heterogeneous models refer to models with different sizes.}, tailored to the specific resource scale available to each client.

\begin{figure}[t]
  \centering
  \includegraphics[width=1.\columnwidth]{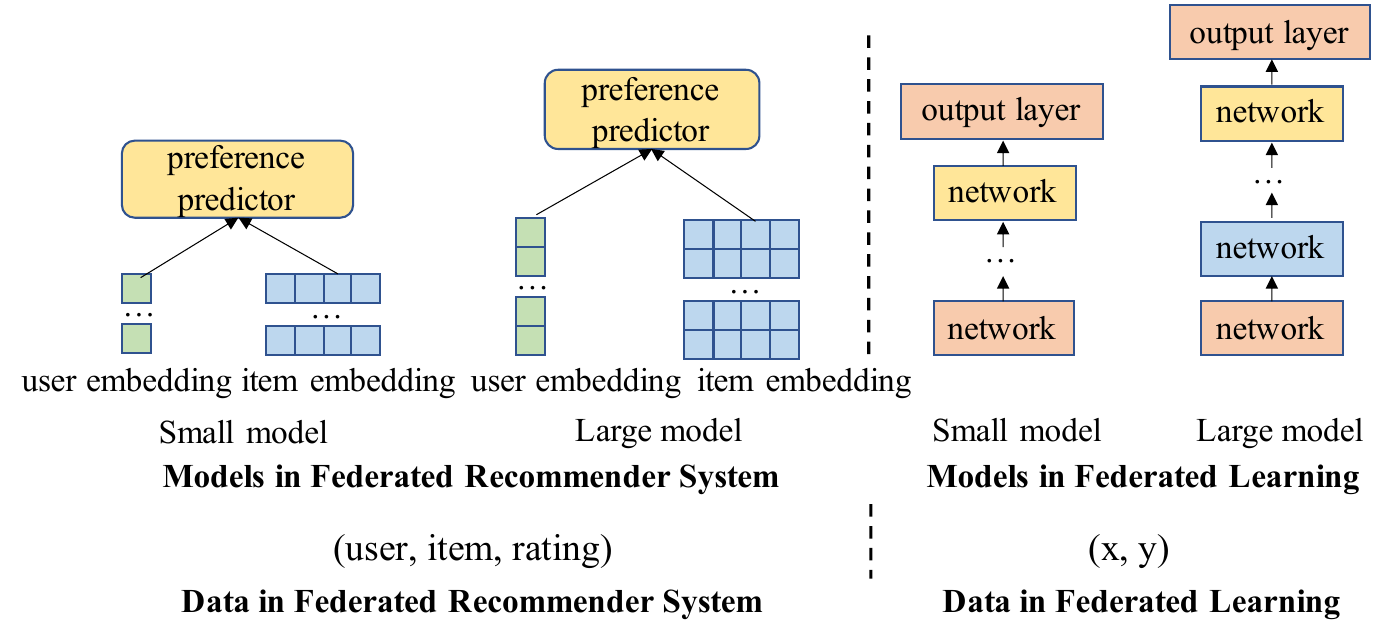} 
  \caption{The difference between federated recommender systems and federated learning in model architecture and data structure.}
  \label{fig_fedrec_fl_diff}
  \end{figure}

In federated learning, several endeavours have been undertaken to achieve model heterogeneity~\cite{ma2022state}.
These methods can generally be classified into two categories.
The first~\cite{chang2019cronus,li2019fedmd,sun2020federated} involves leveraging knowledge distillation~\cite{hinton2015distilling} to exchange knowledge over heterogeneous models via a public reference dataset.
Another solution entails designing new aggregation strategies based on the model architecture, such as channel-wise aggregation~\cite{diao2020heterofl,zhu2022resilient} and layer-wise aggregation~\cite{wang2023flexifed}.
Some works mix elements from both of these solutions to achieve their goals~\cite{liu2022no,kim2023depthfl}.
However, all existing heterogeneous federated learning methods cannot be directly used in FedRecs due to the inherent disparities between general federated learning and FedRecs as displayed in Fig~\ref{fig_fedrec_fl_diff}. 
\begin{itemize}
\item \textbf{Difference in Data Structures}. 
In FedRecs, a data sample is formed as $(user, item, rating)$ which is associated with specific users. Consequently, constructing a public reference dataset to facilitate knowledge distillation while ensuring privacy protection becomes impractical.
\item \textbf{Difference in Model Architecture}. In FedRecs, the item embedding table plays a dominant role in determining the size of the recommendation model. The distinction between small and large models primarily lies in the dimensions of the item embedding. However, the model size in federated learning is mainly related to the depth of layers and the widths of channels.
\end{itemize}

In this paper, we propose a novel federated recommendation framework named HeteFedRec, which aims to facilitate collaborative learning across clients with heterogeneous models. 
Since item embeddings significantly influence the entire recommendation model and user embeddings are typically private parameters not shared among clients, HeteFedRec primarily focuses on heterogeneous item embeddings. The conventional idea is to employ the padding method to aggregate item embeddings of different sizes. However, this aggregation strategy suffers from a severe mismatch problem because parameters at the same position but in different embedding tables may represent different latent factors and meanings. 
To address this issue, we propose unified dual-task learning, ensuring that submatrices of large embedding tables and small embedding tables share the same objective/task. 
Furthermore, we design a novel decorrelation regularization to prevent large item embeddings from degrading to smaller ones. 
Lastly, a relation-based ensemble distillation is proposed to effectively merge knowledge from embeddings of different sizes based on the intuitive idea that the ensembled spatial information derived from various types of item embeddings contains more reliable and useful knowledge.
To showcase the effectiveness, efficiency, and generalization of HeteFedRec, we conduct extensive experiments on three recommendation datasets using two commonly used base recommendation models.

The main contributions of this paper are as follows:
\begin{itemize}
    \item To the best of our knowledge, we are the first to explore model heterogeneity in federated recommender systems.
    \item We propose a heterogeneous model aggregation strategy tailored for FedRecs, which leverages a unified dual-task learning mechanism with dimensional decorrelation regularization to facilitate knowledge sharing among recommender models of different sizes. Furthermore, a relation-based distillation method is proposed to distill knowledge from heterogeneous item embedding tables.
    \item Extensive experiments on three real-world recommendation datasets demonstrate that our proposed methods are effective and efficient and outperform all baselines. 
\end{itemize}

\section{Related Work}\label{sec_relatedwork}
In this section, we briefly review the literature on most related topics. Other involved topics such as general recommender systems~\cite{li2021lightweight} and federated learning can be referred to corresponding surveys~\cite{hung2017computing,nguyen2017argument,yu2023self,zheng2023automl,zhang2021survey}.
\subsection{Federated Recommender Systems} 
Due to the privacy-preserving ability, FedRecs have received remarkable attention recently and have been widely deployed in many recommendation scenarios~\cite{imran2023refrs} such as news~\cite{qi2020privacy}, social~\cite{liu2022federated}, and POI recommendation~\cite{long2023model}.
The pioneering work by~\cite{ammad2019federated} introduced the first federated recommendation framework that utilizes federated learning with collaborative filtering models.
Building upon this basic framework, extensive research has been proposed to improve the performance of FedRecs in a short time~\cite{wang2021fast}.
Some works aim to bridge the performance gap between FedRecs and centralized recommender systems by using advanced neural networks~\cite{wu2021fedgnn} and training techniques~\cite{wu2022fedcl}.
Others focus on reducing the training costs associated with FedRecs.
For example, \cite{chen2018federated,muhammad2020fedfast} studies achieving convergence within fewer training epochs, and~\cite{zhang2023lightfr} attempts to reduce communication costs each round.
Besides, some researchers propose personalized federated recommender systems~\cite{luo2022personalized}, that attempt to balance the common-sense knowledge from global models and the personalized information from local updating.
These personalized federated recommender systems aim to address the content heterogeneity in clients' data, which is orthogonal to this work. 

With the notable advancements of FedRecs, the security concerns surrounding these systems have recently gained attention.
Studies such as~\cite{zhang2022pipattack,rong2022fedrecattack,yuan2023manipulating,yuan2023manipulating1} demonstrate that FedRecs are susceptible to manipulation by malicious users who upload poisoned model updates, resulting in unfair and inappropriate recommendations.
\cite{zhang2022comprehensive,yuan2023interaction,yuan2023federated} reveal that FedRecs still suffer potential privacy issues since adversaries can steal attribute information~\cite{zhang2021graph} and the system cannot forget quit clients.

Although significant progress has been made in FedRecs, existing approaches necessitate sharing a global model of the same size among clients, leading to suboptimal performance. To address this concern, this paper takes a pioneering step with HeteFedRec, enabling clients to collaboratively learn global models with varying sizes in FedRecs.

\subsection{Heterogeneous Federated Learning} 
Generally, the heterogeneity research~\cite{gao2022survey} in federated learning can be classified into three scenarios: data heterogeneity, device heterogeneity, and model heterogeneity.
Data heterogeneity research focuses on learning a global model with clients whose training data is non-IID~\cite{sun2021heterogeneous,ma2022state,ye2023heterogeneous}.
These studies attempt to strike a balance between acquiring common-sense knowledge and maintaining local personalization.
For instance, Zhao et al.~\cite{zhao2018federated} addressed this problem by adding a small subset of global data shared among all the clients.
Li et al.~\cite{li2020federated} proposed FedProx, a re-parametrization of FedAvg~\cite{mcmahan2017communication} that includes a regularization term to preserve the local model's personalization.
Hermes~\cite{li2021hermes} performs partial parameter averaging to retain model personalization.
In addition,~\cite{li2021model} incorporates contrastive learning to rectify model representations among individual parties.
On the other hand, research on device heterogeneity considers variations in computing power among different participants~\cite{jiang2022fedmp,wang2022fedadmm}.

Our work is more related to model heterogeneity federated learning.
Traditional federated learning makes an essential assumption that all clients train a model with the same architecture~\cite{zhang2021survey}.
In model heterogeneity, this assumption is relaxed and the main challenge lies in aggregating knowledge from diverse models.
One research line is to design specific aggregation strategies based on target model architecture.
For example, ~\cite{diao2020heterofl,horvath2021fjord} design width-level aggregation for parameters with different channel scales.
~\cite{liu2022no,wang2023flexifed,kim2023depthfl} control model size by adjusting the number of neural layers, and they only aggregate layers in the lower layers, as these layers capture fundamental features.
Based on layer-level aggregation, \cite{jiang2022fedmp,zhu2022resilient} applies neural pruning techniques to create heterogeneous models.
Another research direction involves using knowledge distillation~\cite{hinton2015distilling} to transfer knowledge among heterogeneous models.
~\cite{chang2019cronus,li2019fedmd} construct a public reference dataset that is accessible to all clients, leveraging soft label distillation to aggregate heterogeneous local models.

However, the aforementioned methods for model heterogeneity cannot be directly applied to FedRecs due to the notable distinctions between FedRecs and traditional federated learning. 
In this paper, we take the initial stride toward implementing model heterogeneity in the context of FedRecs.

\subsection{Knowledge Distillation}
Based on the types of knowledge being transferred, distillation can be classified into response-based knowledge distillation, feature-based knowledge distillation, and relation-based knowledge distillation.
Response-based knowledge distillation~\cite{yuan2020revisiting,tang2020understanding} transfers knowledge from a teacher model to a student model by utilizing the outputs of the final layer.
Feature-based distillation, on the other hand, focuses on training a student model using features extracted from intermediate layers, as these layers learn to discriminate specific features~\cite{zagoruyko2016paying}.
Meanwhile, relation-based distillation involves capturing the relationships between feature maps and leveraging this knowledge to train student models~\cite{chen2021distilling,shang2021lipschitz,yang2022mutual}. 
In this paper, we propose a novel relation-based ensemble self-distillation method that allows for the fusion of knowledge among item embeddings of varying sizes.

\section{Preliminaries}\label{sec_preliminaries}

\subsection{Federated Recommendation Settings}\label{sec_basic_fedrec}
Let $\mathcal{U}=\{u_{i}\}_{i=1}^{\left|\mathcal{U}\right|}$ and $\mathcal{V}=\{v_{j}\}_{j=1}^{\left|\mathcal{V}\right|}$ denote the sets of users and items, respectively, while $\left|\mathcal{U}\right|$ and $\left|\mathcal{V}\right|$ represent their sizes. 
In the context of FedRecs, each user $u_{i}$ is a client that possesses a local dataset $\mathcal{D}_{i}$ with the size $\left|\mathcal{D}_{i}\right|$ varying among users as shown in Fig.~\ref{fig_distribution}.
Note that to protect user privacy, the private dataset $\mathcal{D}_{i}$ will not be shared with any other participants.
$\mathcal{D}_{i}$ consists of user-item interactions $(u_{i}, v_{j}, r_{ij})$, where $r_{ij}$ is implicit feedback in this paper.
$r_{ij}=1$ indicates $u_{i}$ has interacted with item $v_{j}$, and $r_{ij}=0$ means no interaction between $u_{i}$ and $v_{j}$, i.e., $v_{j}$ is a negative sample.
The goal of FedRecs is to predict $\hat{r}_{i*}$ between $u_{i}$ and non-interacted items and subsequently recommend the top-K ones with the highest predicted scores.
To achieve this, FedRecs are trained to optimize the following formula:
\begin{equation}
    \label{eq_simple_objective}
    \begin{aligned}
      \mathop{argmin}\limits_{\{\mathbf{u}_{1,\dots,\left|\mathcal{U}\right|}, \mathbf{V}, \mathbf{\Theta}\}}\sum\limits_{u_{i}\in \mathcal{U}} \mathcal{L}(\mathcal{F}(\mathbf{u}_{i}, \mathbf{V}, \mathbf{\Theta})|\mathcal{D}_{i})\\
    \end{aligned}
  \end{equation}
where $\mathbf{u}_{i}$ is a user embedding vector, $\mathbf{V}$ is a $\left|\mathcal{V}\right| \times N$ item embedding matrix which is the most memory-intensive component in FedRecs.
$N$ is the embedding size (i.e., the number of dimensions). The user embedding $\mathbf{u}_{i}$ is an $1 \times N$ vector.
$\mathbf{\Theta}$ is other trainable parameters such as the parameters of feedforward layers.
To optimize E.q.~\ref{eq_simple_objective} with user privacy protection, FedRecs set $\mathbf{u}_{i}$ as private parameters which will only store in corresponding local devices and will not be accessed by any other participants. 
In contrast, $\mathbf{V}$ and $\mathbf{\Theta}$ are used as public parameters to achieve collaborative learning.
$\mathcal{F}$ is a base recommendation algorithm, in this paper, we tried NCF~\cite{he2017neural} and LightGCN~\cite{he2020lightgcn} which will be introduced in the following subsection.
$\mathcal{L}$ is a loss function, in this paper, following~\cite{zhang2019deep,yuan2023federated,yuan2023interaction,yuan2023manipulating,yuan2023manipulating1}, our loss function is:
\begin{equation}\label{eq_ori_loss}
    \mathcal{L}(\mathbf{u}_{i}\!,\!\mathbf{V}\!,\!\mathbf{\Theta})\!=\!\sum\limits_{(u_{i},\!v_{j},\!r_{ij})\in \mathcal{D}_{i}}\!-r_{ij}\!\log\!\hat{r}_{ij}\!+\!(r_{ij}-1)\log (1-\hat{r}_{ij})
\end{equation}

\begin{figure*}[t]
  \centering
  \includegraphics[width=1.6\columnwidth]{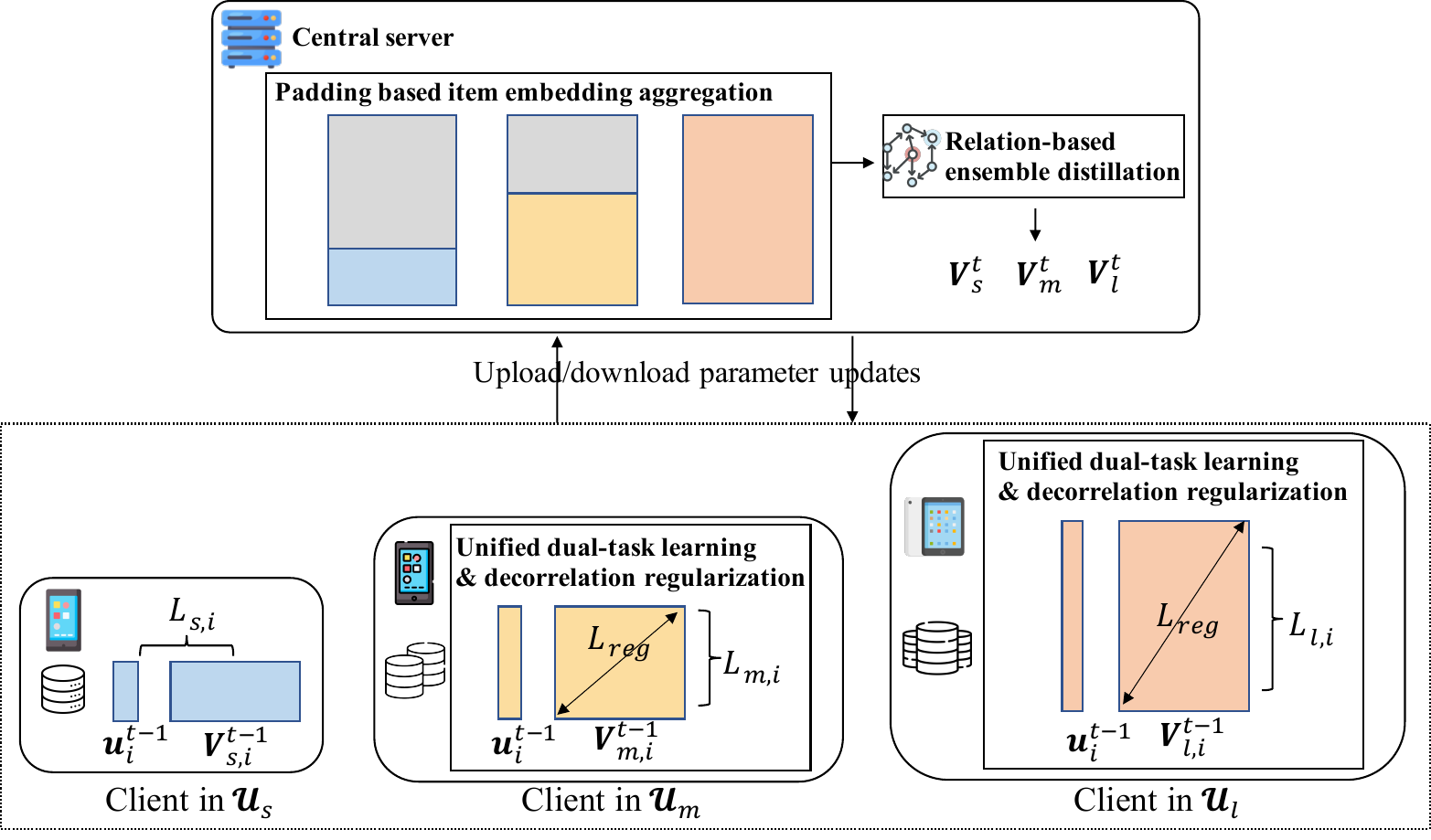} 
  \caption{The framework of HeteFedRec. To facilitate presentation, we omit $\mathbf{\Theta}$ in this figure.}
  \label{fig_item_aggregation}
   \vspace{-20pt}
  \end{figure*}

In FedRecs, a central server coordinates all clients to optimize E.q.~\ref{eq_simple_objective} by transmitting and aggregating the public parameters with the following federated learning protocol.
In the initial stage, the central server initializes all public parameters $\mathbf{V}^{0}$ and $\mathbf{\Theta}^{0}$, and each client independently initializes its private parameters $\mathbf{u}_{i}^{0}$.
After that, the following steps are iteratively executed until model convergence. 
At time step $t$, the central server first selects a group of clients $\mathcal{U}^{t-1}$ and distributes public parameters $\mathbf{V}^{t-1}$ and $\mathbf{\Theta}^{t-1}$ to these clients. 
The selected clients integrate the received public parameters with their private parameters $\mathbf{u}_{i}^{t-1}$, forming local recommenders, which are then trained to optimize E.q.~\ref{eq_ori_loss} on the respective local datasets. 
After several epochs of local training, the client $u_{i}$ sends the updates of public parameters $\nabla \mathbf{V}_{i}^{t-1}$ and $\nabla \mathbf{\Theta}_{i}^{t-1}$ to the central server and concurrently updates its private parameters locally.
\begin{equation}\label{eq_u_update}
  \mathbf{u}_{i}^{t} = \mathbf{u}_{i}^{t-1} - lr\nabla \mathbf{u}_{i}^{t-1}
\end{equation}
where $lr$ is the learning rate.
The server employs specific strategies~\cite{mcmahan2017communication} to aggregate the updates of public parameters and use them to achieve collaborative learning:
\begin{equation}\label{eq_aggregate}
  \begin{aligned}
    &\mathbf{V}^{t} = \mathbf{V}^{t-1} - lr\sum\limits_{u_{i}\in \mathcal{U}^{t-1}} \nabla \mathbf{V}^{t-1}_{i}\\
    &\mathbf{\Theta}^{t} = \mathbf{\Theta}^{t-1} - lr\sum\limits_{u_{i}\in \mathcal{U}^{t-1}} \nabla \mathbf{\Theta}^{t-1}_{i}  
  \end{aligned}
\end{equation}

The above most widely used FedRec framework relies on an important assumption that all clients possess ample resources, such as sufficient training data, to support the training of recommendation models of the same sizes. 
However, as depicted in Fig.~\ref{fig_distribution}, users' data size suffers substantial variance.
Consequently, the basic FedRec framework can only train a recommender system that attains suboptimal performance.

\subsection{Base Recommendation Models}
The basic FedRec framework is compatible with the majority of deep learning-based recommendation models.
Among these models, NCF~\cite{he2017neural} and LightGCN~\cite{he2020lightgcn} are two most widely used techniques.
Hence, to demonstrate the generalization of our proposed methods, we utilize these two recommendation models as base models in this paper.

Neural collaborative filtering~\cite{he2017neural} (NCF) extends the method of collaborative filtering by utilizing several layers of feedforward network (FNN) to learn the complex patterns from user-item interaction data:
\begin{equation}
    \label{eq_NCF}
    \begin{aligned}
      \hat{r}_{ij} = \sigma(FFN([\mathbf{u}_{i}, \mathbf{v}_{j}]))
    \end{aligned}
  \end{equation}
where $FFN(\cdot)$ is a feedforward network, and its parameters are included in $\mathbf{\Theta}$, $[\cdot]$ denotes concatenation operation.

In LightGCN~\cite{he2020lightgcn}, users and items are treated as distinct nodes and a bipartite graph is constructed based on user-item interactions.
Subsequently, a LightGCN propagation is applied on the graph to compute user and item embeddings.
To ensure privacy, the propagation is only used in user's local graph.
Finally, these user and item embeddings are used to predict users' preference scores via E.q.~\ref{eq_NCF}.

\section{Methodology: HeteFedRec}\label{sec_methodology}
As mentioned in Section~\ref{sec_preliminaries}, the commonly used basic FedRec framework necessitates clients to share public parameters of the same sizes, a.k.a. homogeneous FedRecs. 
However, real-world scenarios often involve clients with varying data sizes. For clients with limited data, their model updates may be unreliable and hinder the convergence of the global recommender system. Conversely, clients with abundant data would prefer to train larger models to obtain more accurate recommendations, benefiting from the stronger representation capabilities of larger models.

Based on the homogeneous FedRec framework, there are two naive solutions to address the aforementioned dilemma.
One is to exclude the updates of public parameters from clients with limited training data and only aggregate updates from clients with sufficient data. However, as illustrated in Fig.~\ref{fig_distribution}, this approach would result in the exclusion of a significant portion of clients, severely compromising the recommendation performance. Another intuitive method is to opt for smaller models that are suitable for the majority of clients. Nevertheless, these smaller models have limited representation capabilities, thereby leading to unsatisfactory recommendation performance.

In this paper, we introduce HeteFedRec, the first heterogeneous federated recommender system. 
HeteFedRec allows for the customization of the recommendation model size for each client based on their available resources.
Clients with limited data can train smaller models, while clients with abundant data can still benefit from the services provided by larger models. 
In the following subsections, we will first introduce the fundamental settings of our HeteFedRec, which builds upon the basic FedRec framework. 
Subsequently, we will present the technical details of HeteFedRec, including (1) a heterogeneous recommendation model aggregation strategy that contains a unified dual-task learning and a dimensional decorrelation regularization and (2) a relation-based ensemble self-distillation.
Fig.~\ref{fig_item_aggregation} presents an overview of HeteFedRec.

\subsection{Settings of HeteFedRec}
In HeteFedRec, the size of a client's recommendation model can be customized based on its specific situation.
In this paper, following~\cite{liu2022no}, without loss of generality, we categorize clients into three groups based on the scale of user-item interactions: small clients $\mathcal{U}_{s}$, medium clients $\mathcal{U}_{m}$, and large clients $\mathcal{U}_{l}$, to convenient the introduction of our proposed methods.
Each group is assigned corresponding public parameters $\{\mathbf{V}, \mathbf{\Theta}\}_{s,m,l}$.
Here, $\mathbf{V}_{s}, \mathbf{V}_{m}, \mathbf{V}_{l}$ are item embeddings of sizes $\left|\mathcal{V}\right|\times N_{s}$, $\left|\mathcal{V}\right|\times N_{m}$, $\left|\mathcal{V}\right|\times N_{l}$, respectively, with the relationship $N_{s}<N_{m}<N_{l}$.
The size of $\mathbf{\Theta}$ depends on the dimension length of $\mathbf{V}$ as it primarily includes parameters in feedforward layers.
Note that the settings of division and dimension sizes are hyper-parameters and can be adjusted according to specific situations.

In general, item embedding table $\mathbf{V}$ dominates the number of parameters in the entire recommender system.
With these settings, users in $\mathcal{U}_{s}$ will have smaller models, while users in $\mathcal{U}_{l}$ will have larger models.
Under this heterogeneous circumstance, FedRecs are trained to optimize the following objective:
\begin{equation}
    \label{eq_heterogeneous_objective}
    \begin{aligned}
      \mathop{argmin}\limits_{\{\mathbf{u}_{1\dots\left|\mathcal{U}\right|}, \{\mathbf{V}, \mathbf{\Theta}\}_{s,m,l}\}}\!\sum\limits_{a\in \{s,m,l\}}\!\sum\limits_{u_{i}\in \mathcal{U}_{a}}\!\mathcal{L}(\mathcal{F}(\mathbf{u}_{i}\!,\!\mathbf{V}_{a}\!,\!\mathbf{\Theta}_{a})|\mathcal{D}_{i})\\
    \end{aligned}
  \end{equation}

One straightforward approach to optimize E.q.~\ref{eq_heterogeneous_objective} is using a clustered-aggregation method.
In this approach, the central server aggregates clients' model updates separately based on their respective model sizes. However, this method falls short of achieving optimal performance due to two main reasons:
(1) the knowledge cannot be effectively shared across different groups, hindering the ability to leverage collaborative information;
(2) since each training sample can only update the corresponding item embedding vector, the item embedding table becomes data-hungry, and it may not be well-trained within each isolated group, leading to suboptimal performance.

Another method is to aggregate these heterogeneous models based on some observations, which has proven effective in general federated learning.
For example, HeteroFL~\cite{diao2020heterofl} directly conducts parameter-averaging over heterogeneous models based on the observation that deep and wide neural networks can drop a tremendous number of parameters per layer.
\cite{liu2022no,wang2023flexifed} utilize layer-wise aggregation, focusing on the lower layers that retain more pre-trained and fundamental knowledge. 
However, all these preconditions are infeasible in the context of FedRecs, rendering these methods inapplicable.
In light of this, HeteFedRec introduces a novel heterogeneous aggregation tailored for federated recommender systems.

\subsection{Heterogeneous Recommendation Model Aggregation}

In FedRecs, the private parameters, specifically user embeddings, are updated locally and do not communicate with other clients. 
As a result, the update of these private parameters can still adhere to the protocol outlined in the basic FedRec framework. 
Consequently, HeteFedRec primarily focuses on the knowledge aggregation of the public parameters, namely $\{\mathbf{V}, \mathbf{\Theta}\}_{s,m,l}$. 
Among these parameters, the item embeddings $\{\mathbf{V}\}_{s,m,l}$ play a particularly crucial role as they dominate the number of parameters in a recommender system. 
In what follows, we will first introduce the aggregation strategy design for item embeddings and then provide a brief overview of the aggregation of $\mathbf{\Theta}$.

\textbf{Padding based Heterogeneous Item Embedding Aggregation.}
At epoch $t$, for the uploaded item embedding updates, HeteFedRec's central server first pads the smaller item embedding updates to align them with the largest one using the following function:
\begin{equation}
  \label{eq_padding}
  \begin{aligned}
    &\nabla_{p} \mathbf{V} \leftarrow padding(\nabla \mathbf{V}|N_{l})
  \end{aligned}
\end{equation}
where $padding(\cdot|N_{l})$ is a padding function that transforms the matrix updates $\nabla\mathbf{V}$ to $\nabla_{p}\mathbf{V}$ of size $\left|\mathcal{V}\right|\times N_{l}$ by filling $\mathbf{0}$ vectors.
After padding, the central server performs the summation operation on all received item embedding updates and the aggregated updates for each type of embedding are as follows:
\begin{equation}
  \label{eq_item_aggregation1}
  \begin{aligned}
    \nabla \mathbf{V}_{agg}^{t-1} = \sum_{u_{i}\in\mathcal{U}_{s}^{t-1}}\!&\nabla_{p}\mathbf{V}_{s,i}^{t-1}\!+\!\sum_{u_{i}\in\mathcal{U}_{m}^{t-1}}\!\nabla_{p}\mathbf{V}_{m,i}^{t-1}\!+\!\sum_{u_{i}\in\mathcal{U}_{l}^{t-1}}\!\nabla\mathbf{V}_{l,i}^{t-1}\\
    &\nabla\mathbf{V}_{s}^{t-1} = \nabla \mathbf{V}_{agg_{[:N_{s}]}}^{t-1}\\
    &\nabla\mathbf{V}_{m}^{t-1} = \nabla \mathbf{V}_{agg_{[:N_{m}]}}^{t-1}\\
    &\nabla\mathbf{V}_{l}^{t-1} = \nabla \mathbf{V}_{agg}^{t-1}
  \end{aligned}
\end{equation}
where $\mathcal{U}_{*}^{t-1}$ is the selected clients in the group $\mathcal{U}_{*}$ at epoch $t$. $\nabla\mathbf{V}_{*_{[:N_{x}]}}$ refers to a submatrix of size $\left|\mathcal{V}\right|\times N_{x}$.

Then, the aggregated updates are used to modify corresponding item embeddings:
\begin{equation}\label{eq_hete_aggregate}
  \begin{aligned}
    &\mathbf{V}_{s}^{t} = \mathbf{V}_{s}^{t-1} - lr\nabla \mathbf{V}^{t-1}_{s}\\
    &\mathbf{V}_{m}^{t} = \mathbf{V}_{m}^{t-1} - lr\nabla \mathbf{V}^{t-1}_{m}\\
    &\mathbf{V}_{l}^{t} = \mathbf{V}_{l}^{t-1} - lr\nabla \mathbf{V}^{t-1}_{l}\\
  \end{aligned}
\end{equation}
Note that we set the same learning rate $lr$ for all types of item embeddings since we do not want to introduce too many hyperparameters and we find this setting is effective enough.

According to the aggregation, when HeteFedRec initializes $\mathbf{V}_{s}^{0}$, $\mathbf{V}_{m_{[:N_{s}]}}^{0}$ and $\mathbf{V}_{l_{[:N_{s}]}}^{0}$, $\mathbf{V}_{m_{[N_{s}:N_{m}]}}^{0}$ and $\mathbf{V}_{l_{[N_{s}:N_{m}]}}^{0}$ from the same point respectively, we can get following relationship:
\begin{equation}\label{eq_hete_relationship}
  \begin{aligned}
    &\mathbf{V}_{s}^{t} = \mathbf{V}_{m_{[:N_{s}]}}^{t} = \mathbf{V}_{l_{[:N_{s}]}}^{t}\\
    &\mathbf{V}_{m}^{t} = \mathbf{V}_{l_{[:N_{m}]}}^{t}\\
  \end{aligned}
\end{equation}

\textbf{Unified Dual-task Learning.}
However, the above naive aggregation cannot effectively train usable FedRecs because it suffers from severe mismatch problems of updates.
Specifically, during clients' local training, the updates of $\nabla\mathbf{V}_{s,i}^{t-1}$, $\nabla\mathbf{V}_{m,i}^{t-1}$, and $\nabla\mathbf{V}_{l,i}^{t-1}$ are calculated by optimizing the loss function E.q.~\ref{eq_ori_loss}.
Hence, as a part of these updates, the meaning of the subparts of updates, e.g., $\nabla\mathbf{V}_{m_{[:N_{s}]},i}^{t-1}$, $\nabla\mathbf{V}_{l_{[:N_{s}]},i}^{t-1}$, and $\nabla\mathbf{V}_{l_{[:N_{m}]},i}^{t-1}$, is unclear.
In other words, the subparts of updates, $\nabla\mathbf{V}_{m_{[:N_{s}]},i}^{t-1}$ and $\nabla\mathbf{V}_{l_{[:N_{s}]},i}^{t-1}$, or $\nabla\mathbf{V}_{l_{[:N_{m}]},i}^{t-1}$, are not computed to guide $\mathbf{V}_{s,i}^{t-1}$ or $\mathbf{V}_{m,i}^{t-1}$ to solve the recommendation problem, therefore, it would be difficult to get convergence by aggregating using E.q.~\ref{eq_item_aggregation1}.

\begin{figure}[t]
  \centering
  \includegraphics[width=1.\columnwidth]{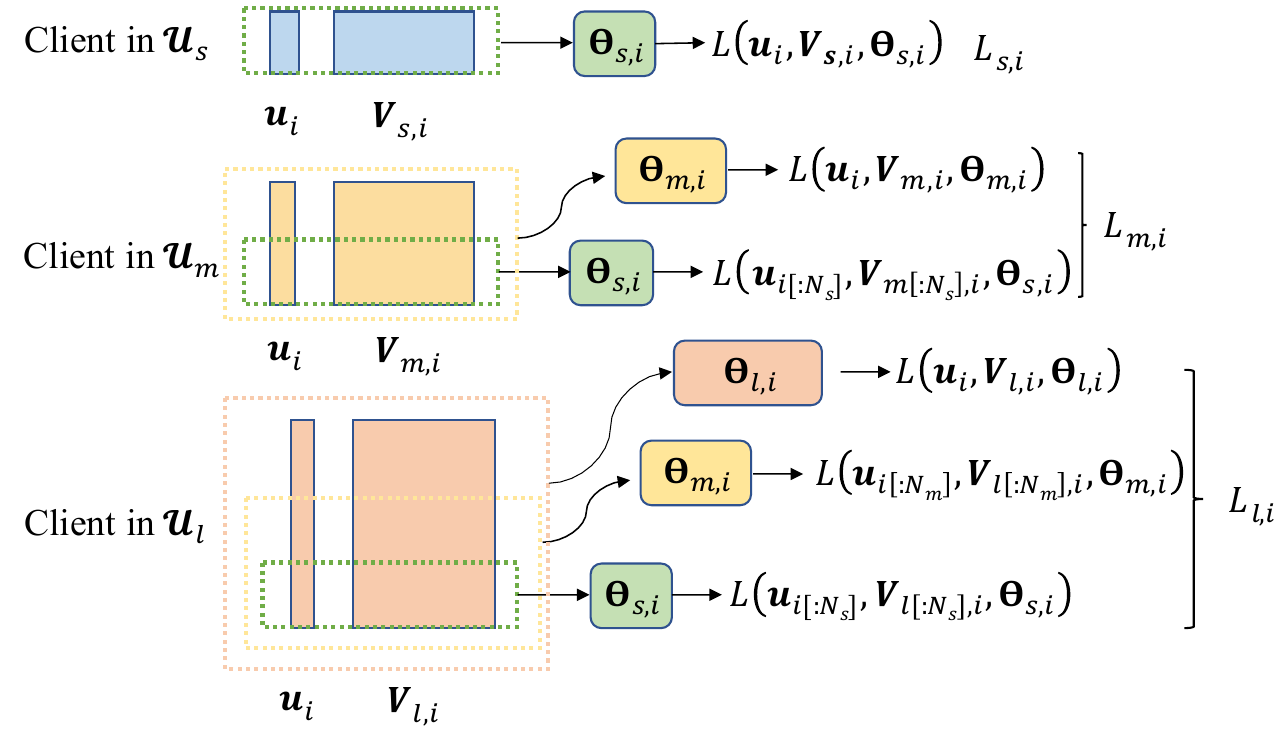} 
  \caption{Overview of unified dual-task learning.}
  \label{fig_dual_task}
   \vspace{-10pt}
  \end{figure}

Inspired by~\cite{kim2023depthfl} that leverages multiple the same objective functions to optimize a deep model's submodel with different depths,
HeteFedRec employs a unified dual-task learning mechanism to ensure submatrices within larger item embeddings share the same objective as small item embeddings. 
Specifically, each user $u_{i}$ will utilize one of the following loss functions depending on the group $\mathcal{U}_{s,m,l}$ it belongs to:
\begin{equation}
  \label{eq_HeteFedRec_obj}
  \begin{aligned}
    & \mathcal{L}_{s,i}=\mathcal{L}(\mathbf{u}_{i}^{t-1},\mathbf{V}_{s,i}^{t-1}, \mathbf{\Theta}_{s,i}^{t-1})\\
    & \mathcal{L}_{m,i}=\mathcal{L}(\mathbf{u}_{i_{[:N_{s}]}}^{t-1},\mathbf{V}_{m_{[:N_{s}]},i}^{t-1}, \mathbf{\Theta}_{s,i}^{t-1}) +  \mathcal{L}(\mathbf{u}_{i}^{t-1},\mathbf{V}_{m,i}^{t-1},\mathbf{\Theta}_{m,i}^{t-1})\\
    & \mathcal{L}_{l,i}\!=\!\mathcal{L}(\mathbf{u}_{i_{[:N_{s}]}}^{t-1},\!\mathbf{V}_{l_{[:N_{s}]},i}^{t-1},\mathbf{\Theta}_{s,i}^{t-1})\!+\!\mathcal{L}(\mathbf{u}_{i_{[:N_{m}]}}^{t-1},\!\mathbf{V}_{l_{[:N_{m}]},i}^{t-1},\mathbf{\Theta}_{m,i}^{t-1})\\
    &\quad\quad\quad+\mathcal{L}(\mathbf{u}_{i}^{t-1},\mathbf{V}_{l,i}^{t-1},\mathbf{\Theta}_{l,i}^{t-1})    
  \end{aligned}
\end{equation}
where $\mathcal{L}(\mathbf{u}_{i}, \mathbf{V}, \mathbf{\Theta})$ is the loss function defined in E.q.~\ref{eq_ori_loss}.
Note that in order to calculate E.q.~\ref{eq_HeteFedRec_obj}, clients in $\mathcal{U}_{m}$ will not only receive $\mathbf{\Theta}_{m}$ but also $\mathbf{\Theta}_{s}$, meanwhile, users in $\mathcal{U}_{l}$ will have all the $\mathbf{\Theta}_{s}$, $\mathbf{\Theta}_{m}$ and $\mathbf{\Theta}_{l}$.
Since $\mathbf{\Theta}$ mainly comprises parameters from feedforward networks with $2$ or $3$ layers, these extra communication costs will be negligible.
Based on the above equation, larger item embeddings' submatrices updates, such as $\nabla\mathbf{V}_{m_{[:N_{s}]},i}^{t-1}$, $\nabla\mathbf{V}_{l_{[:N_{s}]},i}^{t-1}$, and $\nabla\mathbf{V}_{l_{[:N_{m}]},i}^{t-1}$, can be directly aggregated to optimize smaller item embeddings since these updates are trained to minimize the same recommendation loss function.
Fig.~\ref{fig_dual_task} illustrates the unified dual-task learning.

\textbf{Dimensional Decorrelation Regularization.}
Although E.q.~\ref{eq_HeteFedRec_obj} fixes the updates mismatch problem, it may introduce a new challenge known as dimensional collapse for larger embeddings $\mathbf{V}_{m}$ and $\mathbf{V}_{l}$.
Specifically, by solely optimizing the submatrices $\mathbf{V}_{l_{[:N_{s}]},i}$ and $\mathbf{V}_{m_{[:N_{s}]},i}$, all terms in $\mathcal{L}_{l}$ or $\mathcal{L}_{m}$ are reduced. 
This phenomenon causes the item representation of larger models to be primarily confined to the low-dimensional space, rendering the remaining parameters $\mathbf{V}_{m_{[N_{s}:]},i}$ and $\mathbf{V}_{l_{[N_{s}:]},i}$ redundant. 
Consequently, HeteFedRec degrades to homogeneous FedRecs that solely rely on $\mathbf{V}_{s}$, thus diminishing the benefits of model heterogeneity.

To maintain the importance of each dimension in a matrix, one natural way is to retain their singular values of the covariance matrix. 
Therefore, to prevent $\mathbf{V}_{m,i}$ and $\mathbf{V}_{l,i}$ collapsing to small embedding $\mathbf{V}_{s,i}^{t}$, we can add the regularization term $\mathcal{L}_{singular}(\mathbf{V}_{m,i}^{t-1})$ and $\mathcal{L}_{singular}(\mathbf{V}_{l,i}^{t-1})$ in E.q.~\ref{eq_HeteFedRec_obj} to penalize the variance among each dimension's singular value:
\begin{equation} 
  \label{eq_singular_regularization}
  \begin{aligned}
    & \mathcal{L}_{singular}(\mathbf{V})=\frac{1}{N}\sum\limits_{i=1}^{N}(\lambda_{i}-\frac{1}{N}\sum\limits_{j=1}^{N}\lambda_{j})^{2}
  \end{aligned}
\end{equation}
where $\lambda$ is the singular value of $\mathbf{V}$'s covariance matrix.
However, the computation of singular values is expensive. 
~\cite{hua2021feature,shi2022towards} have demonstrated that instead of penalizing the variance of singular values, regularizing the Frobenius norm of the correlation matrix can achieve the same effects.
Thus, in this paper, we employ the following decorrelation regularization term to ensure that $\mathbf{V}_{l_{[N_{s}:]},i}^{t}$ or $\mathbf{V}_{m_{[N_{s}:]},i}^{t}$ encode unique knowledge:
\begin{equation}
  \label{eq_HeteFedRec_obj_with_regularization}
  \begin{aligned}
    & \mathcal{L}_{reg}(\mathbf{V})=\frac{1}{N}\left\| \mathrm{corr}(\frac{\mathbf{V} - \bar{\mathbf{V}}}{\sqrt{\mathrm{var}(\mathbf{V})}})\right\|_{F}\\
  \end{aligned}
\end{equation}
$\left\|\cdot\right\|_{F}$ is Frobenius norm. $\mathrm{corr}(\cdot)$ calculates the correlation matrix. $\bar{\mathbf{V}}$ is the columns' mean of $\mathbf{V}$ by default.
$\mathrm{var}(\cdot)$ computes the variance of a matrix. 

Then, the loss function for clients in groups $\mathcal{U}_{m}$ and $\mathcal{U}_{l}$ is transformed from E.q.~\ref{eq_HeteFedRec_obj} to:
\begin{equation}
  \label{eq_final_loss}
  \begin{aligned}
    & \mathcal{L}_{m,i}^{'}=\mathcal{L}_{m,i} + \alpha \mathcal{L}_{reg}(\mathbf{V}_{m,i}^{t-1})\\
    & \mathcal{L}_{l,i}^{'}= \mathcal{L}_{l,i} + \alpha \mathcal{L}_{reg}(\mathbf{V}_{l,i}^{t-1})
  \end{aligned}
\end{equation}
where $\alpha$ is the factor that controls the importance of the regularization term. 
We adopt the same $\alpha$ for both medium and large item embeddings to minimize the number of hyperparameters. Through experiments, we have observed that this setting proves to be sufficiently effective.

Now, after clients finish local training with the loss function E.q.~\ref{eq_final_loss}, HeteFedRec's central server can conduct heterogeneous aggregation for item embeddings among different models using E.q.~\ref{eq_item_aggregation1}.

\textbf{$\{\mathbf{\Theta}\}_{s,m,l}$ Aggregation.} 
For $\mathbf{\Theta}$, we simply aggregate them with the same size:
\begin{equation}
  \label{eq_theta_aggregation}
  \begin{aligned}
    &\mathbf{\Theta}_{s}^{t} = \mathbf{\Theta}_{s}^{t-1} - lr\sum_{u_{i}\in\mathcal{U}_{s}^{t-1} \cup \mathcal{U}_{m}^{t-1} \cup \mathcal{U}_{l}^{t-1}}\nabla\mathbf{\Theta}_{s,i}^{t-1}\\
    &\mathbf{\Theta}_{m}^{t} =  \mathbf{\Theta}_{m}^{t-1} - lr\sum_{u_{i}\in\mathcal{U}_{m}^{t-1} \cup \mathcal{U}_{l}^{t-1}}\nabla\mathbf{\Theta}_{m,i}^{t-1}\\
    &\mathbf{\Theta}_{l}^{t} = \mathbf{\Theta}_{l}^{t-1} - lr\sum_{u_{i}\in\mathcal{U}_{l}^{t-1}}\nabla\mathbf{\Theta}_{l,i}^{t-1}
  \end{aligned}
\end{equation}

\begin{algorithm}[!ht]
  \renewcommand{\algorithmicrequire}{\textbf{Input:}}
  \renewcommand{\algorithmicensure}{\textbf{Output:}}
  \caption{HeteFedRec: Federated Recommender System with Model Heterogeneity.} \label{alg_hetefedrec}
  \begin{algorithmic}[1]
    \Require global epoch $T$; local epoch $L$; learning rate $lr$, \dots
    \Ensure public parameters $\{\mathbf{V,\mathbf{\Theta}}\}_{s,m,l}$, local client embeddings $\mathbf{u}_{i}|_{i \in \mathcal{U}}$
    \State Initialize public parameters $\{\mathbf{V}^{0},\mathbf{\Theta}^{0}\}_{s,m,l}$
    \For {each round t =1, ..., $T$}
      \State sample a fraction of clients $\mathcal{U}^{t-1}$ from $\mathcal{U}$
        \For{$u_{i}\in \mathcal{U}^{t-1}$ \textbf{in parallel}} 
        \State // execute on client sides
        \State \Call{ClientTrain}{$u_{i}$}
        \EndFor
      \State // the server executes heterogeneous aggregation
      \State $\{\nabla \mathbf{V}^{t-1}\}_{s,m,l}\leftarrow$ aggregate updates using E.q.~\ref{eq_item_aggregation1}
      \State $\{ \mathbf{V}^{t},\mathbf{\Theta}^{t}\}_{s,m,l}\leftarrow$ update public parameters using E.q.~\ref{eq_hete_aggregate} and E.q.~\ref{eq_theta_aggregation}
      \State // the server executes distillation
      \State $\{\mathbf{V}^{t}\}_{s,m,l}\leftarrow$ update embeddings using E.q.~\ref{eq_distillation_loss}
    \EndFor
    \Function{ClientTrain} {$u_{i}$}
    \State download public parameters from the server
    \If{$u_{i}\in \mathcal{U}_{s}$}
          \State calculate $\nabla \mathbf{u}_{i}^{t-1}$, $\nabla \mathbf{V}_{s,i}^{t-1}$, $\nabla \mathbf{\Theta}_{s,i}^{t-1}$ using $\mathcal{L}_{s,i}$ in E.q.~\ref{eq_HeteFedRec_obj}
          \State upload $\nabla \mathbf{V}_{s,i}^{t-1}$,  $\nabla \mathbf{\Theta}_{s,i}^{t-1}$ to the server
        \ElsIf{$u_{i}\in \mathcal{U}_{m}$}
          \State calculate $\nabla \mathbf{u}_{i}^{t-1}$, $\nabla \mathbf{V}_{m,i}^{t-1}$, $\{\nabla \mathbf{\Theta}_{i}^{t-1}\}_{s,m}$ using $\mathcal{L}_{m,i}^{'}$ in E.q.~\ref{eq_final_loss}
          \State upload $\nabla \mathbf{V}_{m,i}^{t-1}$,  $\{\nabla \mathbf{\Theta}_{i}^{t-1}\}_{s,m}$ to the server
        \Else  $\quad$// $u_{i}\in \mathcal{U}_{l}$
          \State calculate $\nabla \mathbf{u}_{i}^{t-1}$, $\nabla \mathbf{V}_{l,i}^{t-1}$, $\{\nabla \mathbf{\Theta}_{i}^{t-1}\}_{s,m,l}$ using $\mathcal{L}_{l,i}^{'}$ in E.q.~\ref{eq_final_loss}
          \State upload $\nabla \mathbf{V}_{l,i}^{t-1}$,  $\{\nabla \mathbf{\Theta}_{i}^{t-1}\}_{s,m,l}$ to the server
        \EndIf
        \State $\mathbf{u}_{i}^{t}\leftarrow$ update local private parameter using E.q.~\ref{eq_u_update}
  \EndFunction
    \end{algorithmic}
\end{algorithm}

\subsection{Relation-based Ensemble Self Knowledge Distillation}
Knowledge distillation has been used to facilitate knowledge sharing among heterogeneous models in federated learning~\cite{gao2022survey}.
However, these works~\cite{liu2022no,chang2019cronus,li2019fedmd} require the construction of publicly available reference datasets.
In the context of FedRecs, where data samples are sensitive and specific to users, these knowledge distillation methods cannot be directly applied to heterogeneous FedRecs.

In this paper, we propose a novel relation-based ensemble self-distillation mechanism to further enhance the performance of heterogeneous FedRecs, eliminating the need for constructing a reference dataset.
The basic idea of our method is that, when item embeddings are well-trained from collaborative information, the similarity of items should be consistent among different embedding sizes.
Therefore, we transfer knowledge by harmonizing the spatial information across item embeddings of different sizes, i.e., keep the relative distance relationship of item vectors consistent among item embedding tables.

Specifically, after getting $\{\mathbf{V}^{t}\}_{s,m,l}$ using heterogeneous recommendation model aggregation at epoch $t$, the central server conducts distillation with the following steps.
First, to avoid heavy computation costs, it randomly selects a subset of items $\mathcal{V}_{kd}$ from the whole item set $\mathcal{V}$ as the target distillation items. 
Then, it calculates selected items' distance from each other using $\{\mathbf{V}^{t}\}_{s,m,l}$ and obtains the ensemble distance by averaging:
\begin{equation}
  \label{eq_ensemble_distill}
  \begin{aligned}
    & d_{ens} (\mathcal{V}_{kd}) =\frac{1}{3} \sum\limits_{a\in\{s,m,l\}} d(\mathbf{V}_{a}^{t},\mathcal{V}_{kd})\\
  \end{aligned}
\end{equation}
where $d(\cdot)$ calculates the distance among selected items. In this paper, we use cosine similarity as the distance function.
Finally, for each embedding table in $\{\mathbf{V}^{t}\}_{s,m,l}$, the central server computes distillation loss as follows:
\begin{equation}
  \label{eq_distillation_loss}
  \begin{aligned}
    & \mathcal{L}_{kd}(\mathbf{V})=   \left\| d(\mathbf{V},\mathcal{V}_{kd}) - d_{ens} (\mathcal{V}_{kd})\right\|_{2}^{2}
      \end{aligned}
\end{equation}
Algorithm~\ref{alg_hetefedrec} summarizes HeteFedRec using pseudocode.

\section{Experiments}\label{sec_experiments}
In this section, 
 we conduct experiments to answer the following research questions (RQs):
\begin{itemize}
  \item \textbf{RQ1.} How effective is our HeteFedRec compared to homogeneous and heterogeneous FedRec baselines?
  \item \textbf{RQ2.} How efficient is our HeteFedRec compared to the baselines in terms of model convergence speed and communication cost?
  \item \textbf{RQ3.} How does HeteFedRec benefit from each key component?
  \item \textbf{RQ4.} How does the ratio of client division affect HeteFedRec's performance?
  \item \textbf{RQ5.} How does the model size affect HeteFedRec's performance?
  \item \textbf{RQ6.} How does the value of hyperparameter $\alpha$ affect HeteFedRec's performance? 
\end{itemize}

\subsection{Datasets}
We conduct extensive experiments on three real-world recommendation datasets: MovieLens-1M (ML)~\cite{harper2015movielens}, Anime, and Douban~\cite{zhu2019dtcdr}, covering three different scenarios (movie recommendation, anime recommendation, and book recommendation) to evaluate the performance of our HeteFedRec.
The statistics of these datasets are presented in Table~\ref{tb_dataset}.
``Avg.'' is the average number of user-item interactions.
``$<50\%$'' or ``$<80\%$'' represent 50\% or 80\% of users' interaction numbers that are less than certain values.
In this paper, we use these values to divide clients into $\mathcal{U}_{s,m,l}$ without specific notation.
The ML dataset consists of users' ratings for movies and includes $6,040$ users, $3,706$ items, and $1,000,209$ interactions. 
The Anime dataset contains user preference data crawled from MyAnimeList\footnote{https://myanimelist.net/} and consists of $10,482$ users, $6,888$ items, and $1,265,530$ watching records.
The Douban dataset is a subset from Douban book, comprising $330,268$ interactions between $1,833$ users and $7,397$ items.
Following the settings of recommendation with implicit feedback~\cite{he2017neural,zhang2022pipattack,yuan2023interaction}, we binarize all datasets' user ratings, transforming all ratings to $r_{ij} = 1$, and negative instances are sampled with a ratio of $1:4$. 
For each dataset, $80\%$ of data and $20\%$ of data are used as training and test set. 
When a client is selected for training, $10\%$ of its training data will be used as the validation set to guide the local training.
Note that to protect data privacy, following~\cite{he2017neural,zhang2022pipattack,yuan2023interaction},  a user is a client, and users can only use their own data during the training process.
\begin{table}[!ht]
  \centering
  \setlength\tabcolsep{4.pt}
  \caption{Statistics of recommendation datasets}\label{tb_dataset}
  
  \begin{tabular}{lccccccc}
  \hline
  \textbf{Dataset}  & \multicolumn{1}{l}{\textbf{Users}} & \multicolumn{1}{l}{\textbf{Items}} & \multicolumn{1}{l}{\textbf{Interactions}} & \multicolumn{1}{l}{\textbf{Avg.}} & \multicolumn{1}{l}{$\mathbf{<50\%}$} & \multicolumn{1}{l}{$\mathbf{<80\%}$} \\ \hline
  ML    & 6,040                                  & 3,706                                 & 1,000,209                                     & 165                                      & 77                               & 203\\
  Anime        & 10,482                                 & 6,888                                 & 1,265,530                                     & 120                                       & 69                       &    150    \\
  Douban  & 1,833                               & 7,397                                & 330,268                                     & 180                                        & 115                               &  244 \\ 
  \hline
  \end{tabular}
  \end{table}

\subsection{Evaluation Metrics} 
To evaluate the effectiveness of the proposed heterogeneous federated recommendation framework, we employ the widely adopted metrics~\cite{he2020lightgcn,ren2023joint} Recall at Rank 20 (Recall@20) and Normalized Discounted Cumulative Gain at Rank 20 (NDCG@20) to evaluate the recommender system's performance.
Recall measures the average probability of relevant items being successfully recommended to users, while NDCG takes into account the position of relevant items in the recommendation list. 
These metrics provide a comprehensive evaluation of the recommender system's performance.

\subsection{Baselines}
Since this is the first work on federated recommender systems with heterogeneous model sizes, we construct the following baselines for comparison:
\begin{itemize}
  \item \textbf{All Small.} This baseline deploys small recommendation models to all clients without considering their data amounts. While this approach avoids the dilemma of weak clients not being able to support larger models, it limits the overall ability of FedRecs.
  \item \textbf{All Large.} This baseline deploys large recommendation models to all clients without considering their data amounts. However, since clients with small amounts of data cannot effectively support the training of large models, their updates can even negatively impact the overall performance of FedRecs.
  \item \textbf{All Large/Exclusive.} This baseline deploys large recommendation models to all clients but only aggregates updates from clients with sufficient data.
  \item \textbf{Standalone.} This baseline assigns heterogeneous recommendation models to clients based on their data amounts. Each client independently trains its model without collaboration or knowledge sharing with other clients.
  \item \textbf{Clustered FedRec.} Following the clustered scheme in federated learning~\cite{ghosh2020efficient,meng2022improving}, we construct Clustered FedRec. It aggregates the same type of model.
  \item \textbf{Directly Aggregate.} This method directly aggregates heterogeneous recommendation model updates using E.q.~\ref{eq_item_aggregation1}. However, since these updates are not guided by the unified dual-task learning mechanism, it results in a significant updates mismatch problem.
\end{itemize}
In general, the first three baselines, ``All Small'', ``All Large'', and ``All Large/Exclusive'', are homogeneous FedRecs, where the recommendation models are either all small or all large, regardless of the client's data amounts.
On the other hand, the last three baselines, ``Standalone'', ``Clustered FedRec'', and ``Directly Aggregate'', are heterogeneous FedRecs inspired by corresponding federated learning schemes. 
These heterogeneous FedRecs consider the varying data amounts of clients and aim to address the challenges posed by resource heterogeneity in the federated recommender system setting.

\begin{table*}[!htbp]
  \centering
  \caption{The comparison of the overall performance of HeteFedRec and baselines. The best values on each dataset are bold. Underline indicates the best values in homogeneous methods.}\label{tb_overall_perform}
  \begin{tabular}{l|l|l|cc|cc|cc}
  \hline
                                    &     &                                                                          & \multicolumn{2}{c|}{\textbf{ML}}     & \multicolumn{2}{c|}{\textbf{Anime}}               & \multicolumn{2}{c}{\textbf{Douban}}     \\ \hline
                                    &\textbf{Type}     & \textbf{Methods}                                                         & \textbf{Recall}    & \textbf{NDCG}      & \textbf{Recall}    & \textbf{NDCG}                & \textbf{Recall}    & \textbf{NDCG}      \\ \hline
  \multirow{7}{*}{\textbf{Fed-NCF}} &\multirow{3}{*}{\textbf{Homogeneous}}     & \textbf{All Small}                                                       & 0.02203          & \underline{0.04328}          & \underline{0.04301}          & \underline{0.04962}                    & \underline{0.00759}          & \underline{0.01087}          \\
                                    &     & \textbf{All Large}                                                       & \underline{0.02558}          & 0.04028          & 0.02727          & 0.04442                    & 0.00726          & 0.00878          \\
                                    &     & \textbf{\begin{tabular}[c]{@{}l@{}}All Large/Exclusive\end{tabular}}   & 0.00956          & 0.01753          & 0.01199          & 0.02458                    & 0.00702          & 0.00856           \\\cline{2-9}
                                    &\multirow{4}{*}{\textbf{Heterogeneous}}     & \textbf{Standalone}                                                      & 0.00615          & 0.01108          & 0.00279          & 0.00411                    & 0.00209          & 0.00295                \\ 
                                    &     & \textbf{Clustered FedRec}                                                & 0.01712          & 0.02235          & 0.01508          & 0.01581                    & 0.00248          & 0.00501                \\
                                    &     & \textbf{Directly Aggregate}                                              & 0.01177          & 0.02207          & 0.01903          & 0.03151                    & 0.00247          & 0.00502                  \\
                                    &     & \textbf{HeteFedRec(Ours)}                                                & \textbf{0.02662} & \textbf{0.04781} & \textbf{0.05855} & \textbf{0.05655} & \textbf{0.01101} & \textbf{0.01290} \\ \hline
  \multirow{7}{*}{\textbf{Fed-LightGCN}}& \multirow{3}{*}{\textbf{Homogeneous}} & \textbf{All Small}                & 0.02251                  & \underline{0.04232}                  & \underline{0.02924}          & \underline{0.04824}                    & \underline{0.00350}          & \underline{0.00530}          \\
                                    &     & \textbf{All Large}                                                       & \underline{0.02301}                  & 0.04197                  & 0.02825          & 0.04788                    & 0.00234          & 0.00378          \\
                                    &     & \textbf{\begin{tabular}[c]{@{}l@{}}All Large/Exclusive\end{tabular}}   & 0.00924                  & 0.01891                  & 0.01702          & 0.01467                    & 0.00215          & 0.00363                 \\\cline{2-9}
                                    &\multirow{4}{*}{\textbf{Heterogeneous}}     & \textbf{Standalone}                                                      & 0.00605                  & 0.01085                  & 0.00278          & 0.00411                    & 0.00190          & 0.00263                  \\
                                    &     & \textbf{Clustered FedRec}                                                & 0.01483                  & 0.02633                  & 0.01443          & 0.01379                    & 0.00259          & 0.00480                  \\
                                    &     & \textbf{Directly Aggregate}                                              & 0.01454                  & 0.02657                  & 0.01450          & 0.01437                    & 0.00257          & 0.00479                  \\
                                    &     & \textbf{HeteFedRec(Ours)}                                                & \textbf{0.02434} & \textbf{0.04313} & \textbf{0.03306} & \textbf{0.05177}           & \textbf{0.00393} & \textbf{0.00639} \\ \hline
  \end{tabular}
  \end{table*}

  \begin{figure*}[t]
    \centering
    \includegraphics[width=1.6\columnwidth]{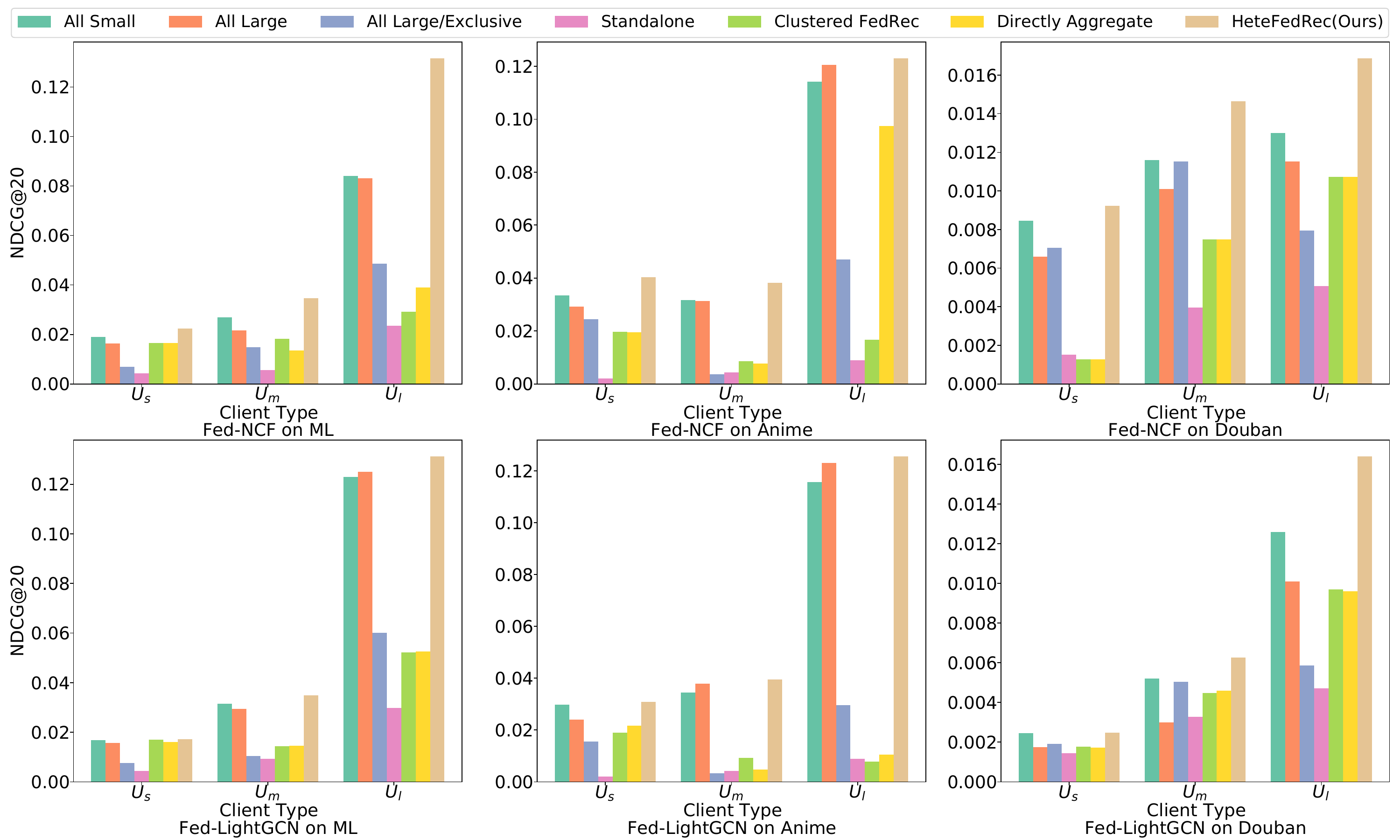} 
    \caption{Detailed performance of ``All Small'', ``All Large'', and ``HeteFedRec(Ours)'' in different user groups.}
    \label{fig_fedclientgroup}
     \vspace{-10pt}
    \end{figure*}
\subsection{Parameter Settings}\label{sec_hyperprametersetting}
On ML and Anime datasets, the dimensions $N_{s}$, $N_{m}$, and $N_{l}$ are $8$, $16$, $32$ for both Fed-NCF and Fed-LightGCN.
For the Douban, $N_{s}$, $N_{m}$, and $N_{l}$ are set to $32$, $64$, and $128$ for both Fed-NCF and Fed-LightGCN as users in Douban have a higher average number of interactions than the other datasets.
The proportion of $\mathcal{U}_{s}$, $\mathcal{U}_{m}$, and $\mathcal{U}_{l}$ is $5:3:2$ without specific noting.
Both Fed-NCF and Fed-LightGCN models utilize three feedforward layers with $[2\times N_{*}, 8, 8]$ dimensions for different types of models. 
The layer of LightGCN propagation is $1$.
Adam~\cite{kingma2014adam} with $0.001$ learning rate is adopted as the optimizer.
At the beginning of an epoch, the server shuffles the queue of clients. 
Then, at each epoch, there are several rounds for the central server to traverse the client queue.
During each round, the central server selects $256$ users for training.

\subsection{The Effectiveness of HeteFedRec (RQ1)}\label{sec_main_exp}
  \begin{table}[!htbp]
    \setlength\tabcolsep{4.pt}
    \caption{The cost of one-time transmitting between clients and the central server.}\label{tb_communication_costs}
    \begin{tabular}{c|ccc}
    \hline
    \textbf{Client Type}   & \textbf{All Small} & \textbf{All Large} & \textbf{HeteFedRec} \\ \hline
    $\mathcal{U}_{s}$ &   size($\mathbf{V}_{s}+\mathbf{\Theta}_{s}$)                 &   size($\mathbf{V}_{l}+\mathbf{\Theta}_{l}$)                 &   size($\mathbf{V}_{s}+\mathbf{\Theta}_{s}$)                   \\
    $\mathcal{U}_{m}$ &   size($\mathbf{V}_{s}+\mathbf{\Theta}_{s}$)                 &   size($\mathbf{V}_{l}+\mathbf{\Theta}_{l}$)                 &   size($\mathbf{V}_{m}+\{\mathbf{\Theta}\}_{s,m}$)                   \\
    $\mathcal{U}_{l}$ &   size($\mathbf{V}_{s}+\mathbf{\Theta}_{s}$)                 &   size($\mathbf{V}_{l}+\mathbf{\Theta}_{l}$)                 &   size($\mathbf{V}_{l}+\{\mathbf{\Theta}\}_{s,m,l}$)                   \\ \hline
    \end{tabular}
    \end{table}

Table~\ref{tb_overall_perform} presents the comparison of HeteFedRec with six baselines.
Among homogeneous baselines, we can see that ``All Small'' achieves better performance than ``All Large''. 
This phenomenon supports our argument that clients with insufficient data cannot support large models' training therefore even with stronger models, ``All Large'' cannot achieve better performance.
Besides, we can observe a significant performance drop of the method ``All Large/Exclusive'', since it excludes the updates from $\mathcal{U}_{s}$ which represent a large proportion of the entire client set.
In heterogeneous FedRecs, ``Standalone'' exhibits the poorest performance due to the lack of collaboration among clients.
The clustered scheme, which has been used in federated learning to address model heterogeneity, proves ineffective for training a usable recommender system in ``Clustered FedRec''.
This could be attributed to the fact that collaborative information is crucial in recommender systems, and the clustered scheme hinders collaborative learning among clients.
The  ``Directly Aggregate'' method attempts to fuse knowledge between heterogeneous models but suffers from severe updates mismatch problems. 
Essentially, ``Directly Aggregate'' can be viewed as HeteFedRec removed unified dual-task learning, dimensional decorrelation regularization, and knowledge distillation components.
In baselines, all heterogeneous FedRecs fail to train a FedRec with comparable performance to homogeneous baselines.
As the first heterogeneous method tailored for FedRecs, our HeteFedRec outperforms all heterogeneous and homogeneous baselines on all datasets using both Fed-NCF and Fed-LightGCN by a significant margin, demonstrating the effectiveness of our proposed methods.

\begin{table*}[!htbp]
  \centering
  \caption{Ablation Study. ``RESKD'' is short for ``Relation-based Ensemble Self Knowledge Distillation'', ``DDR'' is abbreviation for ``Dimensional Decorrelation Regularization'', and ``UDL'' indicates ``Unified Dual-task Learning''. }\label{tb_ablation}
  \begin{tabular}{l|l|cc|cc|cc}
    \hline
                                           &                            & \multicolumn{2}{c|}{\textbf{ML}}    & \multicolumn{2}{c|}{\textbf{Anime}} & \multicolumn{2}{c}{\textbf{Douban}} \\ \hline
                                           &            & \textbf{Recall}  & \textbf{NDCG}    & \textbf{Recall}  & \textbf{NDCG}    & \textbf{Recall}  & \textbf{NDCG}    \\ \hline
    \multirow{4}{*}{\textbf{Fed-NCF}}                       & \textbf{HeteFedRec}        & \textbf{0.02662} & \textbf{0.04781} & \textbf{0.05855} & \textbf{0.05655} & \textbf{0.01101} & \textbf{0.01290} \\
                                           & \textbf{- RESKD}         & 0.02620          & 0.04636          & 0.05696          & 0.05504          & 0.00956          & 0.01232        \\
                                           & \textbf{- RESKD,DDR}     & 0.02390          & 0.04332          & 0.05766          & 0.05355          & 0.00797          & 0.01167        \\
                                           & \textbf{- RESKD,DDR,UDL} & 0.01177          & 0.02207          & 0.01903          & 0.03151          & 0.00247          & 0.00502          \\ \hline
    \multirow{4}{*}{\textbf{Fed-LightGCN}} & \textbf{HeteFedRec}        & \textbf{0.02434}          & \textbf{0.04313}          & \textbf{0.03306}          & \textbf{0.05177}          & \textbf{0.00393}          & \textbf{0.00639}          \\
                                           & \textbf{- RESKD}         & 0.02332          & 0.04307          & 0.03282          & 0.05072          & 0.00372          & 0.00618        \\
                                           & \textbf{- RESKD,DDR}     & 0.02231          & 0.04225          & 0.03129          & 0.04812          & 0.00351          & 0.00563        \\
                                           & \textbf{- RESKD,DDR,UDL} & 0.01454          & 0.02657          & 0.01450          & 0.01437          & 0.00257          & 0.00479          \\ \hline
    \end{tabular}
  \end{table*}

\begin{table}[!htbp]
  \centering
  \caption{The variance of singular values in the covariance matrix of the largest item embedding $\mathbf{V}_{l}$. A higher value implies a more severe dimensional collapse.}\label{tb_singular_variance}
  \begin{tabular}{l|l|ccc}
  \hline
                                         &               & \textbf{ML}     & \textbf{Anime}  & \textbf{Douban} \\ \hline
  \multirow{2}{*}{\textbf{Fed-NCF}}      & \textbf{- DDR} & 0.4573          & 0.9190          & 0.0523          \\
                                         & \textbf{+ DDR} & \textbf{0.0974} & \textbf{0.0838} & \textbf{0.0167} \\ \hline
  \multirow{2}{*}{\textbf{Fed-LightGCN}} & \textbf{- DDR} & 0.0459          & 0.0421          & 0.0348          \\
                                         & \textbf{+ DDR} & \textbf{0.0208} & \textbf{0.0240} & \textbf{0.0171} \\ \hline
  \end{tabular}
  \end{table}

  \begin{figure}[t]
    \centering
    \includegraphics[width=1.\columnwidth]{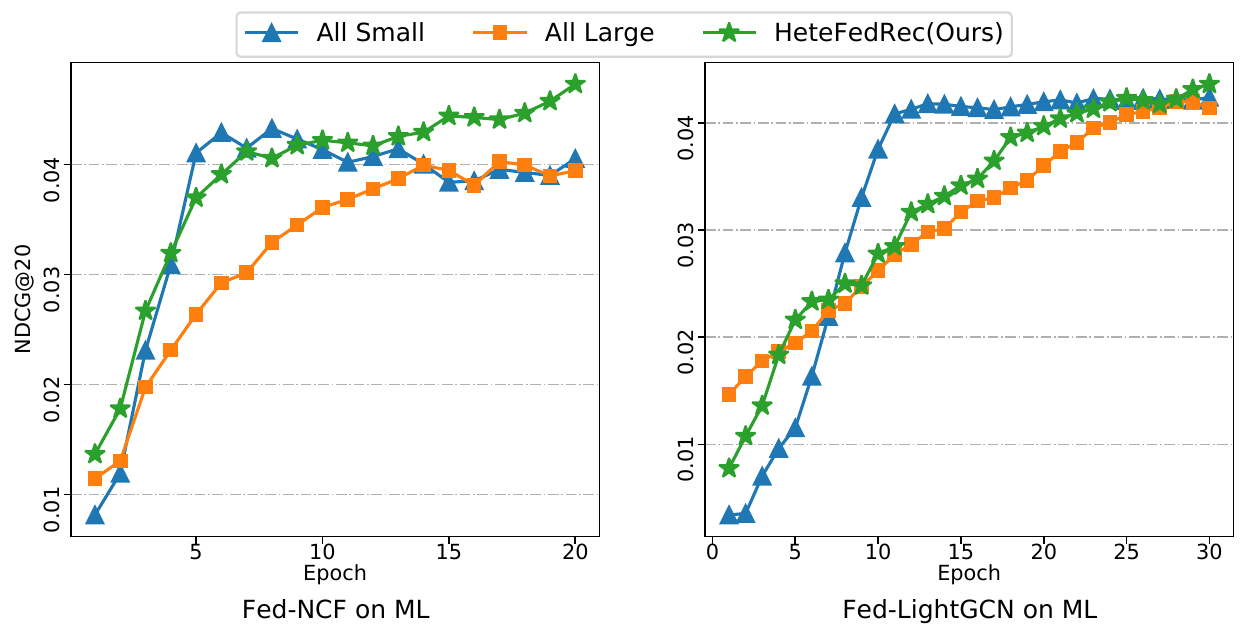}
    \caption{The performance trend during the training process for Fed-NCF and Fed-LightGCN on ML. A similar trend can be seen in Anime and Douban.}
    \label{fig_convergence}
     \vspace{-10pt}
    \end{figure}
  
    \begin{table*}[!htbp]
      \centering
      \caption{The performance of HeteFedRec under different client group division. ``x:y:z'' represents $\frac{x}{x+y+z}$, $\frac{y}{x+y+z}$, and $\frac{z}{x+y+z}$ proportions of users are classified into $\mathcal{U}_{s}, \mathcal{U}_{m},$ and $\mathcal{U}_{l}$ according to interaction amounts, respectively. From 5:3:2 to 2:3:5, more and more clients are assigned larger models. The best values on each dataset are bold.}\label{tb_division}
      \begin{tabular}{l|l|l|ccccc}
        \hline
                                               & \textbf{Datasets}                & \textbf{Metrics} & \textbf{All Small} & \textbf{5:3:2} & \textbf{1:1:1} & \textbf{2:3:5}       & \textbf{All Large} \\ \hline
        \multirow{6}{*}{\textbf{Fed-NCF}}      & \multirow{2}{*}{\textbf{ML}}     & \textbf{Recall}  & 0.02203            & \textbf{0.02662}        & 0.02619      & 0.02576            & 0.02558            \\
                                               &                                  & \textbf{NDCG}    & 0.04328            & \textbf{0.04781}        & 0.04362      & 0.04197            & 0.04028            \\ \cline{2-8} 
                                               & \multirow{2}{*}{\textbf{Anime}}  & \textbf{Recall}  & 0.04301            & \textbf{0.05855}        & 0.05792      & 0.05602            & 0.02727            \\
                                               &                                  & \textbf{NDCG}    & 0.04962            & \textbf{0.05655}        & 0.05540      & 0.05431            & 0.04442            \\ \cline{2-8} 
                                               & \multirow{2}{*}{\textbf{Douban}} & \textbf{Recall}  & 0.00759            & \textbf{0.01101}        & 0.00819      & 0.00756            & 0.00726            \\
                                               &                                  & \textbf{NDCG}    & 0.01087            & \textbf{0.01290}        & 0.00996      & 0.00893            & 0.00878            \\ \hline
        \multirow{6}{*}{\textbf{Fed-LightGCN}} & \multirow{2}{*}{\textbf{ML}}     & \textbf{Recall}  & 0.02251            & \textbf{0.02434}        & 0.02335      & 0.02296            & 0.02301            \\
                                               &                                  & \textbf{NDCG}    & 0.04232            & \textbf{0.04313}        & 0.04233      & 0.04204            & 0.04197            \\ \cline{2-8} 
                                               & \multirow{2}{*}{\textbf{Anime}}  & \textbf{Recall}  & 0.02924            & \textbf{0.03306}        & 0.03233      & 0.03064            & 0.02825            \\
                                               &                                  & \textbf{NDCG}    & 0.04824            & \textbf{0.05177}        & 0.04838      & 0.04804            & 0.04788            \\ \cline{2-8} 
                                               & \multirow{2}{*}{\textbf{Douban}} & \textbf{Recall}  & 0.00350            & \textbf{0.00393}        & 0.00360      & 0.00333            & 0.00234            \\
                                               &                                  & \textbf{NDCG}    & 0.00530            & \textbf{0.00639}        & 0.00528      & 0.00483            & 0.00378            \\ \hline
        \end{tabular}
      \end{table*}

To gain further insight into the performance gap, we analyze the NDCG scores in detailed user groups as shown in Fig.~\ref{fig_fedclientgroup}.
We focus on analyzing the performance of ``All Small'', ``All Large'', and HeteFedRec, as these two homogeneous baselines achieve performances closest to our HeteFedRec.
Overall, the performance of all three methods follows a consistent trend: NDCG scores are higher in $\mathcal{U}_{l}$ and $\mathcal{U}_{m}$ compared to $\mathcal{U}_{s}$.
This is because that $\mathcal{U}_{l}$ and $\mathcal{U}_{m}$ have more training data available compared to $\mathcal{U}_{s}$.
Specifically, on ML and Anime datasets, ``All Small'' outperforms ``All Large'' in $\mathcal{U}_{s}$ for both Fed-NCF and Fed-LightGCN. 
However, in $\mathcal{U}_{l}$, the performance of ``All Small'' is generally worse than ``All Large''. 
This intriguing observation suggests that clients in $\mathcal{U}_{s}$ struggle to train large models effectively, while clients in $\mathcal{U}_{l}$ benefit from the larger models with stronger learning abilities, aligning with our motivation to introduce model heterogeneity in FedRecs
On the Douban dataset, ``All Small'' consistently outperforms ``All Large'' in all user groups. 
This may be attributed to the large model size (i.e., $128$ dimension size) of the embedding tables, making it challenging for all clients to handle the training.
Our HeteFedRec achieves the best performance across all client groups and scenarios.
This can be attributed to two factors: 
(1) the assignment of models of different sizes based on clients' data amounts, enabling clients to train appropriate models of suitable sizes,
and (2) the effective knowledge transfer among heterogeneous recommendation models through our proposed heterogeneous aggregation strategies, leading to performance improvements in all user groups.

\subsection{The Efficiency of HeteFedRec (RQ2)}
Fig.~\ref{fig_convergence} illustrates the convergence of HeteFedRec compared to ``All Small'' and ``All Large'' as these two baselines achieved comparable performance.
Due to the limitation of space, we only present the results on the ML dataset, similar trends can be observed on other datasets.
Generally, HeteFedRec achieves better performance after about $10$ epochs and $20$ epochs in Fed-NCF and Fed-LightGCN respectively.
``All Small'' is the method that reaches convergence within the fewest epochs because it averagely has the smallest recommendation models and small models are usually easy to get convergence.
The convergence speed for ``All Large'' and our HeteFedRec is similar, but the converged performance of HeteFedRec is much better than these two baselines.
In conclusion, HeteFedRec achieves convergence within a reasonable number of epochs compared to ``All Small'' and ``All Large''.

The communication costs are another important consideration when using a FedRec framework. 
Table~\ref{tb_communication_costs} compares HeteFedRec's communication costs with ``All Small'' and ``All Large'' formally.
The only additional costs incurred by HeteFedRec are size($\mathbf{\Theta}_{s}$) for clients in $\mathcal{U}_{m}$ and size($\mathbf{\Theta}_{s,m}$) for users in $\mathcal{U}_{l}$.
Considering our experimental settings and taking the ML dataset as an example, the parameter numbers of $\mathbf{V}_{s}, \mathbf{V}_{m}$, and $\mathbf{V}_{l}$ are $29648$, $59296$ and $118592$, and the parameters of $\{\mathbf{\Theta}_{s,m,l}\}$ are generally from feedforward layers and their sizes are no more than one hundred.
Therefore, these additional costs are negligible.

\subsection{Ablation Study (RQ3)}\label{sec_ablation}
HeteFedRec comprises three crucial components: a unified dual-task learning to address mismatch problems, a dimensional decorrelation regularization to prevent dimensional collapse, and a relation-based self-knowledge distillation to further merge knowledge among heterogeneous models.
In this part, we investigate the impact of each component.
Specifically, we gradually remove knowledge distillation (RESKD), regularization (DDR), and unified dual-task learning (UDL) and present the experimental results in Table~\ref{tb_ablation}.
When RESKD is removed, there is a slight drop in the performance of HeteFedRec, indicating the effectiveness of RESKD.
We then continue by removing the regularization term and find that, HeteFedRec's performance becomes similar to ``All Small'' in Table~\ref{tb_overall_perform}.
This is because, without the dimensional decorrelation regularization term, HeteFedRec experiences a dimensional collapse problem, resulting in a degradation of its model's representation ability to that of small models.
When all three components are eliminated, HeteFedRec becomes equivalent to ``Directly Aggregate'', and the recommender system's performance is significantly decreased, highlighting the crucial role of our unified dual-task learning mechanism.

To further understand the effectiveness of DDR, we calculate the variance of singular values in the covariance matrices of the largest item embeddings $\mathbf{V}_{l}$, as shown in Table~\ref{tb_singular_variance}.
A smaller variance indicates a more balanced importance among dimensions, suggesting the alleviation of the dimensional collapse problem. 
As displayed in Table~\ref{tb_singular_variance}, after applying DDR, the variance of singular values is reduced in all cases.

\subsection{Impact of Client Division (RQ4)}\label{sec_client_division}
In this part, we thoroughly investigate the influence of client division.
Specifically, we conduct experiments with three different division ratios $5:3:2$, $1:1:1$, and $2:3:5$.
The first division setting is the original one used in the main experiments, which assumes that most clients cannot afford the training of larger models.
The $1:1:1$ ratio is a neutral strategy that evenly distributes clients into the $\mathcal{U}_{s}$, $\mathcal{U}_{m}$, and $\mathcal{U}_{l}$ groups based on the number of users' training data.
The $2:3:5$ ratio is an optimistic division solution that assumes only $20\%$ of clients lack training data and most clients have sufficient data to train stronger recommender systems.
Table~\ref{tb_division} presents the experimental results of HeteFedRec under different client divisions.
We also include ``All Small'' and ``All Large'' in Table~\ref{tb_division} for comparison, as they roughly correspond to the division ratios of $10:0:0$ and $0:0:10$, respectively.
Overall, HeteFedRec with the conservative ratio setting (i.e., $5:3:2$) achieves better performance among these three division strategies.
This result is consistent with the interaction number distribution of ML, Anime, and Douban as indicated in Fig.~\ref{fig_distribution}, where most users have insufficient interaction data.
In addition, as we move from the left to the right in Table~\ref{tb_division} (i.e., $5:3:2$ to ``All Large''), more and more clients are assigned with larger models, but the overall performance continues to deteriorate.
This observation further supports the importance of model size heterogeneity in FedRecs.
It is important to note that in this paper, our focus is on providing a framework to achieve model heterogeneity in FedRecs, while the optimal client division estimation can be explored in future research.

\begin{table}[!htbp]
  \setlength\tabcolsep{4.pt}
  \centering
    \caption{The performance (NDCG@20) of HeteFedRec and baselines under different model size settings on ML. A similar trend can be observed in Anime and Douban. $\{a,b,c\}$ represents the size of $N_{s},N_{m}$ and $N_{l}$. From the left to the right part of the table, model sizes are increased. The best values are bold.}\label{tb_model_size}
  \begin{tabular}{l|l|ccc}
  \hline
                                         &                     & \textbf{\{2,4,8\}} & \textbf{\{8,16,32\}} & \textbf{\{32,64,128\}} \\ \hline
  \multirow{3}{*}{\textbf{Fed-NCF}}      & \textbf{All Small}  & 0.03791            & 0.04328              & 0.04028                \\
                                         & \textbf{All Large}  & \textbf{0.04328}   & 0.04028              & 0.03903                \\
                                         & \textbf{HeteFedRec} & 0.03829            & \textbf{0.04781}     & \textbf{0.04074}       \\ \hline
  \multirow{3}{*}{\textbf{Fed-LightGCN}} & \textbf{All Small}  & 0.03813            & 0.04232              & \textbf{0.04197}                \\
                                         & \textbf{All Large}  & \textbf{0.04232}            & 0.04197              & 0.03901                \\
                                         & \textbf{HeteFedRec} & 0.04017            & \textbf{0.04313}     & 0.04093                \\ \hline
  \end{tabular}
  \end{table}

  \subsection{Impact of Model Size (RQ5)}\label{sec_model_sizes}
Since this paper aims to propose a personalized model sizes federated recommender system, the settings of model sizes are important to be explored.
In the main experiments, we set $\{N_{s}, N_{m}, N_{l}\}$ to $\{8,16,32\}$. In this part, we tried a smaller setting (i.e., $\{2,4,8\}$) and a larger setting (i.e., $\{32,64,128\}$). We only present the experimental results on ML in Table~\ref{tb_model_size} due to space limitation.
A similar conclusion can be obtained from other datasets.
At first, by comparing the homogeneous FedRecs (``All Small'' and ``All Large''), we observed that as the model size increased from $2$ to $128$, the FedRec's performance initially improved and then decreased for both Fed-NCF and Fed-LightGCN. 
For example, in Fed-NCF, when $N_{s}$ is $2$, the NDCG score is $0.03791$, indicating that such a small model cannot capture the complex patterns of user-item interactions.
Then, when the model size increased to $8$, the performance improved to approximately $0.043$.
However, when the model size continued to increase to $32$ and $128$, FedRecs' performance dropped to about $0.040$ NDCG scores.

Comparing HeteFedRec with the homogeneous FedRecs, when the setting is $\{2,4,8\}$, its performance is better than ``All Small'' but is slightly behind ``All Large''.
This is because, in such a range, the models are too small, and simply increasing the model size can yield positive feedback. 
Under the setting of $\{8,16,32\}$, HeteFedRec outperforms both ``All Small'' and ``All Large'' due to its personalized model size assignment.
For $\{32, 64, 128\}$, where the model sizes are too large, ``All Small'' achieves the best performance, but our HeteFedRec still outperforms ``All Large''.
In this paper, our focus is on providing a FedRec framework that enables heterogeneous model sizes. 
The task of finding an optimal model size set can be explored in future work.

\begin{figure}[!htbp]
  \centering
  \includegraphics[width=1.\columnwidth]{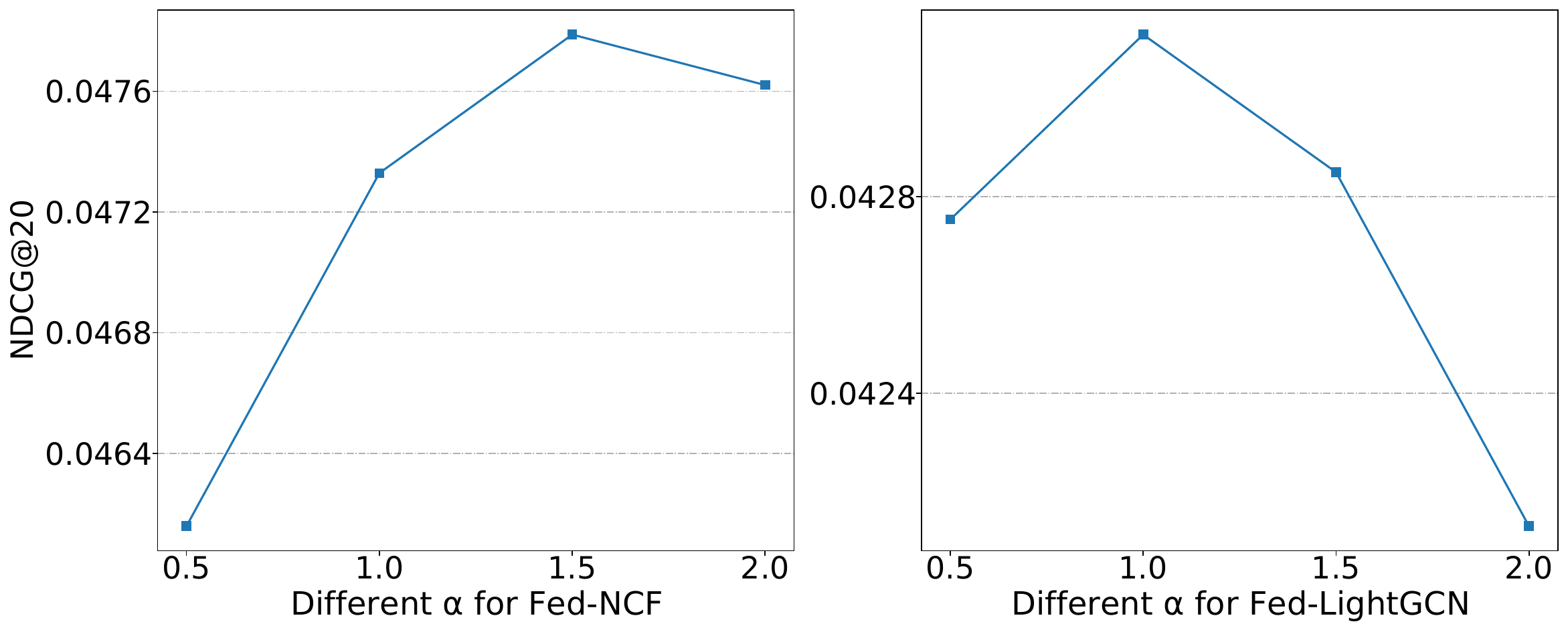}
  \caption{The performance trend w.r.t. the change of $\alpha$ for Fed-NCF and Fed-LightGCN on ML. Similar trends can be seen on the other two datasets.}
  \label{fig_regularization}
   \vspace{-10pt}
  \end{figure}

\subsection{Impact of Regularization Factor $\alpha$ (RQ6)}
In this part, we explore the impacts of the dimensional decorrelation regularization factor $\alpha$.
Fig.~\ref{fig_regularization} shows the performance changes as $\alpha$ increases from $0.5$ to $2.0$ on ML due to space limitation.
In both cases, the trends indicate that HeteFedRec's performance initially increases to a peak point and then decreases with a further increase in $\alpha$.

\section{Conclusion}\label{sec_conclusion}
This paper introduces HeteFedRec, the first federated recommendation framework that allows personalized model sizing for participants.
HeteFedRec leverages unified dual-task learning to facilitate additive aggregation of recommendation models with varying item embedding sizes.
A dimensional decorrelation regularization is employed to prevent dimensional collapse.
Additionally, HeteFedRec incorporates a relation-based knowledge distillation approach to enhance knowledge sharing from diverse recommendation models.
Extensive experiments are conducted on three real-world datasets using two widely used recommender systems. The results demonstrate the effectiveness and generalizability of HeteFedRec. 
However, HeteFedRec still requires the recommendation model to have the same model architecture.
Besides, how to find the optimal solution of client group division and model sizes for each group is also non-trivial as HeteFedRec's performance is very sensitive to these settings.
In future work, we would like to explore address these two problems.

\section*{Acknowledgment}
This work is partially supported by the Australian Research Council under the streams of Future Fellowship (No. FT210100624) and  Discovery Project (DP240101108). 

\bibliographystyle{IEEEtran}
\bibliography{IEEEabrv,IEEEexample}

\end{document}